\documentclass[lettersize,journal]{IEEEtran}
\usepackage{amsmath,amsfonts}
\usepackage{amssymb}
\usepackage{algorithmic}
\usepackage{array}
\usepackage{xcolor}
\usepackage[caption=false,font=normalsize,labelfont=sf,textfont=sf]{subfig}
\usepackage{textcomp}
\usepackage{stfloats}
\usepackage{url}
\usepackage{verbatim}
\usepackage{graphicx}
\usepackage{cite}
\usepackage{algorithm}
\usepackage{bm}
\def\BibTeX{{\rm B\kern-.05em{\sc i\kern-.025em b}\kern-.08em
    T\kern-.1667em\lower.7ex\hbox{E}\kern-.125emX}} 
\usepackage{balance}
\usepackage{amsthm}
\newtheorem{definition}{Definition}
\newtheorem{remark}{Remark}
\newtheorem{proposition}{Proposition}
\newtheorem{assumption}{Assumption}
\newtheorem{lemma}{Lemma}
\newtheorem{theorem}{Theorem}
\newtheorem{corollary}{Corollary}
  
\begin{document}
\title{Information Bottleneck under Perfect Privacy}
\author{Junle Zhong,~\IEEEmembership{}
Mohamad Assaad,~\IEEEmembership{}
\and Sreejith Sreekumar

\thanks{The authors are with Laboratoire des Signaux et Syst\`emes (L2S), CNRS, CentraleSup\'elec, Universit\'e Paris-Saclay, 91190 Gif-sur-Yvette, France (e-mail: junle.zhong@centralesupelec.fr;
mohamad.assaad@centralesupelec.fr; sreejith.sreekumar@centralesupelec.fr).}

\thanks{}}

\maketitle

\begin{abstract}

In this work, we study the information bottleneck under perfect privacy, with particular emphasis on the active-rate regime, where the representation-rate constraint is binding and directly limits the achievable utility. The goal is to construct a representation that preserves utility-relevant information while remaining statistically independent of a sensitive variable. This exact independence requirement introduces an additional constraint beyond the classical rate-relevance tradeoff and must be explicitly incorporated into the optimization. To this end, we develop an alternating direction method of multipliers (ADMM)-based method tailored to the resulting problem structure. Under suitable regularity conditions, we establish global convergence of the generated sequence, characterize its convergence rate through the Kurdyka--Łojasiewicz exponent, and extend the analysis to inexact block updates.

\end{abstract}


\section{Introduction}
The information bottleneck (IB) problem was first introduced in~\cite{tishby2000informationbottleneckmethod} as a fundamental framework for extracting a compressed representation of an observed source while preserving information relevant to a prescribed task. Given random variables $(X,Y)$ with joint distribution $P_{XY}$, the IB problem seeks a stochastic mapping \(P_{U|X}\) that maximizes the relevant information \(I(U;Y)\) subject to a representation-rate constraint \(I(X;U)\le R\). It therefore characterizes the tradeoff between the complexity of a representation and the task-relevant information retained by that representation. This rate--relevance viewpoint has motivated extensive studies of the theoretical properties, operational interpretations, and computational realization of IB problems~\cite{harremoes2007information,asoodeh2020bottleneck}.

However, when the observed data are correlated with sensitive attributes, controlling only representation rate and task relevance is insufficient to prevent the disclosure of private information. The leakage of sensitive information from learned or released representations has therefore become a central concern in privacy-aware representation design. This concern has motivated a broad class of privacy--utility tradeoff formulations, in which a data-release mechanism is designed to balance data utility against privacy protection~\cite{du2012privacy,sankar2013utility,asoodeh2016information}. Within the mutual-information-based setting, a prominent formulation is the privacy funnel (PF), which characterizes the tradeoff between the information retained about a useful variable and the information leaked about a sensitive variable~\cite{makhdoumi2014information}. The most stringent privacy requirement in this framework is perfect privacy, under which the released representation must be statistically independent of the sensitive variable, equivalently satisfying zero information leakage~\cite{calmon2015fundamental,rassouli2021perfect}.

\subsection{Related work}

Information-theoretic privacy-preserving representation design has been studied under several privacy criteria. In the nonzero-leakage regime, the privacy funnel formulates the design of a privacy mechanism as the optimization of a stochastic mapping that generates the released representation while controlling the leakage about the private variable~\cite{makhdoumi2014information}. Related mutual-information-based formulations derive privacy-utility bounds, structural properties, and mechanism-design principles under privacy leakage constraints~\cite{sankar2013utility}. However, since positive leakage is allowed, they do not enforce the exact independence required by perfect privacy.

The zero-leakage regime has also been studied under the notion of perfect privacy, where the released representation is required to be statistically independent of the private variable. Existing works have characterized its fundamental limits, identified feasibility conditions for extracting nontrivial utility, and derived privacy-utility functions in which perfect privacy appears as the zero-leakage case~\cite{calmon2015fundamental,asoodeh2016information}. Structural properties and computational characterizations have also been established for finite alphabet settings, including linear programming formulations for mechanism design~\cite{rassouli2021perfect}. However, they do not explicitly impose a representation-rate constraint \(I(X;U)\le R\), under which the resulting problem no longer admits the same linear programming treatment.

In~\cite{razeghi2023bottlenecks}, the trade-off among complexity, leakage, and utility is studied in a data-driven setting through a unified information-theoretic framework. This formulation is relevant to rate-constrained privacy-preserving representation design, as rate and privacy requirements can be jointly modeled through an information-theoretic Lagrangian. However, the associated algorithmic treatment relies on variational surrogate bounds instead of directly optimizing the original Lagrangian. Consequently, the relation between the surrogate objective and the original problem is not explicitly characterized. In addition, exact equality constraints are not directly handled, and stationarity or convergence guarantees for the resulting nonconvex optimization procedure are not provided. 

The interaction between a representation rate constraint and perfect privacy was studied more directly in~\cite{sreekumar2019optimal}. In particular, when the rate constraint is inactive, the maximum utility under perfect privacy admits a linear programming characterization, which can be viewed as a generalization of~\cite{rassouli2021perfect}. In this regime, the rate budget is sufficiently large and therefore does not affect the optimal privacy--utility operating point. 

Related information-theoretic characterizations were further developed
in~\cite{zamani2023privacy,zamani2026information}. These works studied achievable lower bounds on the maximum achievable utility and established the existence of corresponding representations without and with a rate constraint, respectively. The latter further extends the perfect-privacy setting to bounded leakage \(I(S;U)\leq\epsilon\), with \(\epsilon=0\) recovering perfect privacy. Both works rely on extensions of the Functional Representation Lemma (FRL) and the Strong Functional Representation Lemma (SFRL) to obtain the corresponding achievability results. However, determining the underlying conditional distributions remains challenging.

In many practical settings, the representation rate is limited, making the rate constraint active. Although existing studies provide useful characterizations of the achievable utility under privacy and rate constraints, they do not directly provide a numerical method with convergence guarantees for the perfect-privacy problem in this regime. This motivates the exact constrained formulation and convergence analysis developed in this paper.

From an algorithmic perspective, ADMM provides a natural tool for constrained optimization. Since its early development in~\cite{gabay1976dual,glowinski1975approximation}, ADMM and its variants have been widely used for linearly constrained composite problems. More recently, convergence guarantees have also been established for certain nonconvex and nonsmooth ADMM schemes under appropriate structural assumptions~\cite{hong2016convergence,wang2019global}. In information-theoretic optimization, ADMM was introduced to the information bottleneck problem in~\cite{bayat2019information}, and a provably convergent ADMM-based information bottleneck method was further developed in~\cite{huang2021provably}. A related framework based on Douglas--Rachford splitting and ADMM was proposed in~\cite{huang2022linearly} for a class of Markovian information-theoretic optimization problems, including formulations connected to the information bottleneck and the privacy funnel. Nevertheless, the joint treatment of utility maximization, representation-rate control, and exact perfect-privacy constraints falls outside the direct scope of that formulation. In addition, although its convergence analysis is developed under the structural assumptions of the proposed composite model, how probability-simplex constraints are accommodated in these analyses is not made explicit.

Standard nonconvex ADMM analyses often rely on structural assumptions such as full rank conditions, range inclusion between the constraint matrices, and Lipschitz continuity of the gradient associated with the last updated primal block. In many such analyses, the latter two conditions are used jointly to bound successive dual differences by successive changes in the last primal block. This control is crucial for handling the increase induced by the dual update and establishing descent of the augmented Lagrangian. These assumptions can be restrictive in information theoretic optimization, where the constraint matrices may be rank deficient and probability constraints introduce nonsmooth indicator functions into the block objectives. Moreover, when the augmented Lagrangian itself does not satisfy the required descent property, constructing an appropriate Lyapunov function becomes an additional challenge.

To address these difficulties, methods based on perturbed dual updates have recently been developed to relax some of the restrictive assumptions used in standard nonconvex ADMM analyses~\cite{koshal2011multiuser,Hajinezhad2019,zhang2021online,yang2022proximal,zhou2025perturbed}. The modified dual update provides additional control of both the dual sequence and the feasibility residual, making the perturbation principle particularly relevant to the present formulation with rank deficient constraints. Building on this principle, we develop an ADMM method tailored to the rate constrained perfect privacy problem. Rather than directly applying a general perturbed framework as in~\cite{zhou2025perturbed}, we exploit the specific strongly convex and weakly convex structure of the objective, retain the probability constraints in the block optimality conditions, and introduce proximal regularization only in the block where it is needed to control negative curvature. We further construct a Lyapunov function adapted to the perturbed dual update to establish the required descent property. Furthermore, by exploiting the Kurdyka--{\L}ojasiewicz property of the Lyapunov function, we establish the finite length property and the whole sequence convergence. Moreover, we characterize the convergence rate in terms of the associated K{\L} exponent.

\subsection{Contributions}
The main contributions of this work are summarized as follows.
\begin{itemize}
\item We investigate the active-rate information bottleneck under perfect-privacy constraint. In contrast to weighted leakage formulations, perfect privacy is imposed directly on the representation mechanism, avoiding the need to identify an unknown privacy multiplier capable of recovering the zero-leakage boundary. This leads to a direct solver with convergence guarantees for the active-rate perfect-privacy information bottleneck problem.

\item We propose an ADMM-based computational framework tailored to the resulting nonconvex, nonsmooth, and rank-deficient constrained problem. To the best of our knowledge, this is the first perturbed dual ADMM framework for finite-alphabet information-theoretic optimization. The probability set constraints are explicitly incorporated into both the block updates and the associated optimality conditions.

\item We construct a Lyapunov function to establish convergence of the generated sequence toward a first-order solution of the constrained formulation. By further exploiting the Kurdyka--{\L}ojasiewicz property of the Lyapunov function, we characterize the convergence rates in terms of the associated K{\L} exponent.

\item We further extend the analysis to inexact solutions of the block subproblems under controlled error conditions, and establish sufficient conditions for preserving convergence.
\end{itemize}
\subsection{Organization}
The remainder of this paper is organized as follows. Section~\ref{sec:preliminaries} introduces the notations and preliminary definitions used throughout the paper. Section~\ref{sec:problem} presents the rate-constrained perfect privacy problem and its equivalent reformulation. Section~\ref{sec:proposed} develops the proposed ADMM-based algorithm. Section~\ref{sec:convergence_analysis} establishes the convergence properties of the proposed method. Section~\ref{sec:convergence_rate_analysis} further derives the convergence rates under the Kurdyka--{\L}ojasiewicz framework. 
Section~\ref{sec:numerical_results} presents numerical experiments to illustrate the effectiveness of the proposed method. 
Finally, Section~\ref{sec:conclusion} concludes the paper.

\section{Preliminaries}
\label{sec:preliminaries}

\subsection{Notations}

Throughout this paper, uppercase letters, such as \(X\), denote random variables, which may be vector-valued. Calligraphic letters, such as \(\mathcal X\), denote finite alphabets. The corresponding lowercase letters, such as \(x\), denote realizations of random variables. Bold lowercase letters, such as \(\mathbf x\), denote vector-valued realizations and vectors. Bold uppercase letters, such as \(\mathbf A\), denote matrices and linear operators. The symbols \(\mathbf 0_n\), \(\mathbf 1_n\), and \(\mathbf I_n\) denote the \(n\)-dimensional zero column vector, the \(n\)-dimensional all-ones column vector, and the \(n\times n\) identity matrix, respectively.

For a discrete random variable \(X\) over a finite alphabet \(\mathcal X\), its probability mass function is represented by the column vector
\(
\mathbf p_X \triangleq \bigl[p_X(x)\bigr]_{x\in\mathcal X}.
\)
For two discrete random variables \(X\) and \(Y\), the conditional distribution from \(X\) to \(Y\) is represented by the column-stochastic matrix
\(
\mathbf P_{Y|X}
\triangleq
\bigl[p_{Y|X}(y|x)\bigr]_{y\in\mathcal Y,\;x\in\mathcal X}.
\) 
Thus, \(p_X(x)\) and \(p_{Y|X}(y|x)\) denote scalar probability values, whereas \(\mathbf p_X\) and \(\mathbf P_{Y|X}\) denote the corresponding column vector and matrix representations.

For a positive integer \(n\), \(\mathbb R^n\) denotes the \(n\)-dimensional Euclidean space. For an iterative sequence, \(\mathbf x^k\) denotes the value of \(\mathbf x\) at the \(k\)-th iteration.

For vectors \(\mathbf x,\mathbf y\in\mathbb R^n\), \(\langle \mathbf x,\mathbf y\rangle:=\mathbf x^\top\mathbf y\) denotes the Euclidean inner product and, unless otherwise specified, \(\|\mathbf x\|:=\sqrt{\langle \mathbf x,\mathbf x\rangle}\) denotes the Euclidean norm. For a positive semidefinite matrix \(\mathbf M\), we write \(\|\mathbf x\|_{\mathbf M}^2:=\mathbf x^\top\mathbf M\mathbf x\). The smallest eigenvalue of a matrix \(\mathbf M\) is denoted by \(\lambda_{\min}(\mathbf M)\). \(\operatorname{Im}(\mathbf M)\) denotes the image space of \(\mathbf M\) and \(\otimes\) denotes the Kronecker product. The transpose of \(\mathbf M\) is denoted by \(\mathbf M^\top\). For symmetric matrices \(\mathbf M\) and \(\mathbf N\), the notation \(\mathbf M\succ\mathbf N\) (\(\mathbf M\succeq\mathbf N\)) means that \(\mathbf M-\mathbf N\) is positive definite (positive semidefinite). For a matrix \(\mathbf M\in\mathbb R^{m\times n}\), \(\operatorname{vec}(\mathbf M)\in\mathbb R^{mn}\) denotes the column-wise vectorization of \(\mathbf M\), obtained by stacking its columns from left to right. The differential operator and subdifferential operator are denoted by \(\nabla\) and \(\partial\).

\subsection{Definitions}
For an extended real-valued function \(f:\mathbb R^n\to\mathbb R\cup\{+\infty\}\), \(\operatorname{dom} f:=\{\mathbf x:f(\mathbf x)<+\infty\}\) denotes its effective domain. For a set \(\mathcal C\subseteq\mathbb R^n\), \(\iota_{\mathcal C}\) denotes its indicator function, i.e., \(\iota_{\mathcal C}(\mathbf x)=0\) if \(\mathbf x\in\mathcal C\) and \(\iota_{\mathcal C}(\mathbf x)=+\infty\) otherwise. The distance from \(\mathbf x\) to \(\mathcal C\) is denoted by
\(
\operatorname{dist}(\mathbf x,\mathcal C):=
\inf_{\mathbf y\in\mathcal C}\|\mathbf x-\mathbf y\|.
\)

\begin{definition}[Proper function]
An extended-valued function \(f:\mathbb R^n\to\mathbb R\cup\{+\infty\}\) is proper if
\[
\operatorname{dom} f\ne\emptyset
\quad\text{and}\quad
f(\mathbf x)>-\infty,\ \forall\, \mathbf x\in\mathbb R^n.
\]
\end{definition}

\begin{definition}[Lower semicontinuity]
An extended-valued function \(f:\mathbb R^n\to\mathbb R\cup\{+\infty\}\) is lower semicontinuous at \(\mathbf x\) if
\(
f(\mathbf x)\le \liminf_{\mathbf y\to\mathbf x} f(\mathbf y).
\)
It is lower semicontinuous if it is lower semicontinuous at every \(\mathbf x\in\mathbb R^n\).
\end{definition}

\begin{definition}[Normal cone]
For a closed convex set \(\mathcal C\subseteq\mathbb R^n\), the normal cone to \(\mathcal C\) at \(\mathbf x\in\mathcal C\) is defined as
\[
N_{\mathcal C}(\mathbf x):=
\{\mathbf v\in\mathbb R^n:
\langle \mathbf v,\mathbf y-\mathbf x\rangle\le 0,\ \forall \,\mathbf y\in\mathcal C\}.
\]
\end{definition}
\begin{definition}[Regular and limiting subdifferentials~\cite{rockafellar1998variational}]
Let \(f:\mathbb R^n\to\mathbb R\cup\{+\infty\}\) be finite at \(\mathbf x\). The regular subdifferential of \(f\) at \(\mathbf x\) is defined as
\[
\widehat{\partial} f(\mathbf x)\mspace{-4 mu}:= \mspace{-4 mu}
\left\{\mspace{-2 mu}\mathbf v\in\mathbb R^n:
\liminf_{\substack{\mathbf y\to\mathbf x,\\ \mathbf y\ne\mathbf x}}
\frac{f(\mathbf y)-f(\mathbf x)-\langle \mathbf v,\mathbf y-\mathbf x\rangle}
{\|\mathbf y-\mathbf x\|}\mspace{-2 mu}\ge 0\mspace{-2 mu}
\mspace{-2 mu}\right\}.
\]
The limiting subdifferential of \(f\) at \(\mathbf x\) is defined as
\[
\partial f(\mathbf x):=
\left\{\mathbf v\in\mathbb R^n:
\begin{aligned}
&\exists\,\mathbf x^k\to\mathbf x,\quad
f(\mathbf x^k)\to f(\mathbf x),\\
&\mathbf v^k\in\widehat{\partial}f(\mathbf x^k),
\mathbf v^k\to\mathbf v\ \text{as } k\to+\infty
\end{aligned}
\right\}.
\]
\end{definition}

\begin{remark}
If \(f\) is differentiable at \(\mathbf x\), then \(\widehat{\partial}f(\mathbf x)=\partial f(\mathbf x)=\{\nabla f(\mathbf x)\}\). We use \(\operatorname{crit} f:=\{\mathbf x:\mathbf 0\in\partial f(\mathbf x)\}\) to denote the set of critical points of \(f\).
\end{remark}

\begin{definition}[Lipschitz differentiability]
A differentiable function \(f:\mathbb R^n\to\mathbb R\) is said to be \(L_f\)-Lipschitz differentiable if
\[
\|\nabla f(\mathbf x)-\nabla f(\mathbf y)\|
\le L_f\|\mathbf x-\mathbf y\|,
\qquad \forall \mathbf x,\mathbf y\in\mathbb R^n.
\]
\end{definition}

\begin{definition}[Strong convexity~\cite{rockafellar1998variational}]
A proper function \(f:\mathbb{R}^n\rightarrow\mathbb{R}\cup\{+\infty\}\) is strongly convex with modulus \(\mu>0\) if \(f(\mathbf{x})-\frac{\mu}{2}\|\mathbf{x}\|^2\) is convex.   
\end{definition}
\begin{remark}
\label{rem:strong-convexity}
For a strongly convex function \(f\) with modulus \(\mu>0\), any
\(\mathbf{x}_1,\mathbf{x}_2\in\operatorname{dom}f\), and
\(\mathbf{u}_1\in\partial f(\mathbf{x}_1)\),
\(\mathbf{u}_2\in\partial f(\mathbf{x}_2)\), it holds that
\[
\left\langle
\mathbf{u}_1-\mathbf{u}_2,\,
\mathbf{x}_1-\mathbf{x}_2
\right\rangle
\geq
\mu\|\mathbf{x}_1-\mathbf{x}_2\|^2.
\]
Moreover, the strong convexity of \(f\) also yields
\[
f(\mathbf{x}_1)-f(\mathbf{x}_2)
-
\left\langle
\mathbf{u}_1,\,
\mathbf{x}_1-\mathbf{x}_2
\right\rangle
\leq
-\frac{\mu}{2}
\|\mathbf{x}_1-\mathbf{x}_2\|^2.
\]
\end{remark}

\begin{definition}[Restricted weak convexity~\cite{huang2022linearly}]
\label{def:restricted-weak-convexity}
A function \(f:\mathbb R^n\to[0,+\infty)\) is said to be \(\omega\)-restricted weakly convex with respect to a matrix \(\mathbf A\in\mathbb R^{m\times n}\), where \(\omega\geq0\), if \(f\in C^1\) and
\[
f(\mathbf y)
\geq
f(\mathbf x)
+
\left\langle
\nabla f(\mathbf x),
\mathbf y-\mathbf x
\right\rangle
-
\frac{\omega}{2}
\left\|
\mathbf A(\mathbf y-\mathbf x)
\right\|^2
\]
for all \(\mathbf x,\mathbf y\in\mathbb R^n\).
\end{definition}
\begin{remark}
\label{rem:restricted-weak-convexity}
For an \(\omega\)-restricted weakly convex function \(f\) with respect to \(\mathbf A\), we have
\[
\left\langle
\nabla f(\mathbf x_1)-\nabla f(\mathbf x_2),
\mathbf x_1-\mathbf x_2
\right\rangle
\geq
-\omega
\left\|
\mathbf A(\mathbf x_1-\mathbf x_2)
\right\|^2.
\]
\end{remark}

\begin{definition}[K{\L} property~\cite{Attouch2013}]
Let \(f:\mathbb R^n\to\mathbb R\cup\{+\infty\}\) be proper and lower semicontinuous. The function \(f\) is said to satisfy the Kurdyka-{\L}ojasiewicz property at \(\bar{\mathbf x}\in\operatorname{dom}\partial f\) if there exist \(\eta>0\), a neighborhood \(\mathcal N(\bar{\mathbf x})\) of \(\bar{\mathbf x}\), and a continuous concave function \(\varphi:[0,\eta)\to\mathbb R_+\) such that \(\varphi(0)=0\), \(\varphi\) is continuously differentiable on \((0,\eta)\), \(\varphi'(s)>0\) for all \(s\in(0,\eta)\), and
\[
\varphi'(f(\mathbf x)-f(\bar{\mathbf x}))
\operatorname{dist}(\mathbf 0,\partial f(\mathbf x))
\ge 1
\]
whenever \(\mathbf x\in\mathcal N(\bar{\mathbf x})\) and \(f(\bar{\mathbf x})<f(\mathbf x)<f(\bar{\mathbf x})+\eta\).
\end{definition}

\begin{remark}
If \(f\) satisfies the K{\L} property at every point of \(\operatorname{dom}\partial f\), then \(f\) is called a K{\L} function.
\end{remark}

\section{Problem Formulation}
\label{sec:problem}

Let \(X\), \(Y\), and \(S\) be discrete random variables over finite alphabets \(\mathcal X\), \(\mathcal Y\), and \(\mathcal S\), respectively. The variable \(X\) represents the observed source, \(Y\) the utility-relevant variable, and \(S\) the sensitive variable. A privacy mechanism is specified by a conditional distribution \(\mathbf P_{U|X}\), which maps \(X\) to a released representation \(U\) over a finite alphabet \(\mathcal U\). The cardinality \(|\mathcal U|\) is fixed throughout this paper. Accordingly, the induced joint distribution satisfies the Markov chain
\[
(Y,S)\;\to\;X\;\to\;U .
\]

The utility of the released representation is quantified by the mutual information \(I(U;Y)\), which measures the amount of information that \(U\) preserves about the utility-relevant variable \(Y\). The representation rate is quantified by the mutual information \(I(X;U)\), which measures the amount of information about the observed source \(X\) retained in \(U\). For a prescribed rate budget \(R\ge 0\), the rate constraint is given by \(I(X;U)\le R\). The privacy leakage is measured by \(I(S;U)\). This paper focuses on the perfect privacy regime, namely \(I(S;U)=0\). Let
\[
\mathcal Q
\triangleq
\left\{
\mathbf P_{U|X}
\in
\mathbb R_{\geq 0}^{|\mathcal U|\times|\mathcal X|}
:
\mathbf 1_{|\mathcal U|}^{\top}\mathbf P_{U|X}
=
\mathbf 1_{|\mathcal X|}^{\top}
\right\}
\]
denote the set of all admissible mechanisms. The rate-constrained privacy-utility tradeoff under perfect privacy (RCPP) problem is then formulated as
\begin{equation}
\label{prob:RCPP}
\begin{aligned}
\max_{\mathbf P_{U|X}\in\mathcal Q}\quad
& I(U;Y)\\
\mathrm{s.t.}\quad
& I(X;U)\le R,\\
& I(S;U)=0 .
\end{aligned}
\tag{P1}
\end{equation}

The trivial source independent mechanism is always feasible for~\eqref{prob:RCPP} when \(R\ge 0\), but it yields \(I(U;Y)=0\). Hence, feasibility alone does not guarantee that the problem admits a useful representation. This motivates the following definition of nontrivial perfect privacy.
\begin{definition}[Nontrivial of perfect privacy~\cite{rassouli2021perfect,sreekumar2019optimal}]
Perfect privacy is nontrivial if there exists a mapping \(\mathbf P_{U|X}\) whose output U is statistically dependent on the useful data \(Y\), while being statistically independent of the private data \(S\). 
\end{definition}
\begin{remark}
For finite alphabets, a nontrivial perfect-privacy mechanism exists if and only if
\[
\dim\!\left(
\mathcal N(\mathbf P_{S|X})\cap \mathcal N(\mathbf P_{Y|X})^\perp
\right)\ge 1 .
\]  
\end{remark}
Throughout this work, we assume that this condition holds.

The role of the rate constraint depends on the prescribed value of \(R\).
To make this point precise, we define the maximum utility achievable under perfect privacy as 
\[
J_{\mathrm{PP}}^{\star}
\triangleq
\max_{\substack{\mathbf P_{U|X}\in\mathcal Q:\\ I(S;U)=0}}
I(U;Y),
\]
and define the critical rate
\[
R^{\star}
\triangleq
\min_{\substack{\mathbf P_{U|X}\in\mathcal Q:\\ I(S;U)=0\\
I(U;Y)=J_{\mathrm{PP}}^{\star}}}
I(X;U).
\]
For any \(R\ge R^{\star}\), the rate constraint does not change the maximum perfect-private utility: a perfect-private mechanism achieving \(J_{\mathrm{PP}}^{\star}\) is already feasible under the rate budget. Thus, the problem reduces to the perfect privacy case without an active rate limitation, which can be handled through a linear programming formulation~\cite{sreekumar2019optimal}. By contrast, when \(0<R<R^{\star}\), the rate constraint restricts the set of admissible perfect-private mechanisms and directly affects the attainable utility. This paper focuses on this active-rate regime, where representation-rate control, utility and perfect privacy must be addressed simultaneously.

Following the Lagrange multiplier treatment commonly used in the IB method, we consider the following functional associated with \eqref{prob:RCPP}:
\begin{equation}
\mathcal L_{\mathrm{RCPP}}\big[\mathbf P_{U|X}\big]
\triangleq
I(X;U)-\beta I(U;Y),
\label{eq:rcpp_functional}
\end{equation}
where \(\beta\ge 0\) is a prescribed tradeoff parameter. By the data processing inequality under \(Y\to X\to U\), we have \(I(U;Y)\le I(X;U)\). Therefore, for any \(\beta\le 1\),
\[
I(X;U)-\beta I(U;Y)\ge (1-\beta)I(X;U)\ge 0,
\]
and the value \(0\) is achieved by the trivial representation satisfying \(U\perp X\). Thus, the Lagrange-multiplier formulation admits a trivial global solution for \(\beta\le 1\)~\cite{wu2020learnability}. Let \(\beta^\star\) denote the tradeoff parameter associated with the critical rate \(R^\star\). The active-rate regime \(0<R<R^\star\) therefore corresponds to the nontrivial parameter range \(0<\beta<\beta^\star\). Accordingly, the subsequent analysis is restricted to the nontrivial regime \(1<\beta<\beta^\star\). 

The RCPP problem is then considered in the following form:
\begin{equation}
\begin{aligned}
\min_{\mathbf P_{U|X}\in\mathcal Q}\quad
& I(X;U)-\beta I(U;Y)\\
\mathrm{s.t.}\quad
& I(S;U)=0 .
\end{aligned}
\label{prob:rcpp_lagrangian_problem}
\tag{P2}
\end{equation}
\begin{remark}
In the nondegenerate case, the rate constraint can be represented through an appropriate Lagrange multiplier \(\beta\), as in the standard information bottleneck formulation. Accordingly, we consider the Lagrangian form~\eqref{prob:rcpp_lagrangian_problem} in the subsequent analysis.
\end{remark}
The perfect privacy condition is equivalent to the statistical independence between \(S\) and \(U\), i.e., \(p_{U|S}(u|s)=p_U(u)\) for all \(u\in\mathcal U\) and all \(s\in\mathcal S\) with \(p_S(s)>0\). Under the Markov chain \(S\to X\to U\), we have
\[
\mathbf P_{U|S}=\mathbf P_{U|X}\mathbf P_{X|S},
\qquad
\mathbf p_U=\mathbf P_{U|X}\mathbf p_X .
\]
Thus, the marginal consistency and perfect privacy conditions are respectively given by
\[
\mathbf p_U=\mathbf P_{U|X}\mathbf p_X,
\qquad
\mathbf P_{U|S}=\mathbf p_U\mathbf 1_{|\mathcal S|}^{\top}.
\]

We next express these relations in vector form. Let
\(
\mathbf x\triangleq \mathbf p_U,
\,
\mathbf z\triangleq \operatorname{vec}(\mathbf P_{U|X}).
\)
Then the two relations above can be combined into the single linear constraint
\(
\mathbf A\mathbf x-\mathbf B\mathbf z=\mathbf 0,
\)
where
\[
\mathbf A
\triangleq
\begin{bmatrix}
\mathbf I_{|\mathcal U|}\\
\mathbf 0_{|\mathcal S||\mathcal U|\times|\mathcal U|}
\end{bmatrix},
\qquad
\mathbf B
\triangleq
\begin{bmatrix}
\mathbf p_X^{\mathsf T}\otimes\mathbf I_{|\mathcal U|}\\
\left(
\mathbf P_{X|S}^{\mathsf T}
-
\mathbf 1_{|\mathcal S|}\mathbf p_X^{\mathsf T}
\right)
\otimes\mathbf I_{|\mathcal U|}
\end{bmatrix}.
\]
Let \(\Delta_{\mathcal U}\) denote the probability simplex over \(\mathcal U\). Then the Lagrange-multiplier problem in~\eqref{prob:rcpp_lagrangian_problem} can be written as
\begin{equation}
\begin{aligned}
\min_{\mathbf x,\mathbf z}\quad
& I(X;U)-\beta I(U;Y)\\
\mathrm{s.t.}\quad
& \mathbf A\mathbf x-\mathbf B\mathbf z=\mathbf 0,\\
& \mathbf x\in\Delta_{\mathcal U},\quad
\mathbf z\in\mathcal Q .
\end{aligned}
\label{eq:rcpp_constrained_reformulation}
\tag{P3}
\end{equation}

Based on the constrained reformulation in \eqref{eq:rcpp_constrained_reformulation}, we further rewrite the information objective in a composite form. Since \(\mathbf x\) represents the marginal distribution of \(U\), the entropy term \(H(U)\) depends only on \(\mathbf x\). The remaining conditional entropy terms are evaluated according to the distribution induced by the release mechanism represented by \(\mathbf z\). For a prescribed parameter \(1<\beta<\beta^\star\), we have 
\[
I(X;U)-\beta I(U;Y)
=
(1-\beta)H(U)-H(U|X)+\beta H(U|Y).
\]

In a manner similar to the entropy decomposition used for the IB objective
in~\cite{huang2021provably}, define
\[
\begin{aligned}
F^{0}(\mathbf x)&\triangleq (1-\beta)H(U),&
H^{0}(\mathbf z)&\triangleq -H(U|X)+\beta H(U|Y),\\
F^{1}(\mathbf x)&\triangleq \iota_{\Delta_{\mathcal U}}(\mathbf x),&
H^{1}(\mathbf z)&\triangleq \iota_{\mathcal Q}(\mathbf z).
\end{aligned}
\]
Let
\[
F(\mathbf x)\triangleq F^{0}(\mathbf x)+F^{1}(\mathbf x),
\qquad
H(\mathbf z)\triangleq H^{0}(\mathbf z)+H^{1}(\mathbf z).
\]
The probability constraints \(\mathbf x\in\Delta_{\mathcal U}\) and
\(\mathbf z\in\mathcal Q\) are therefore incorporated into the nonsmooth
parts of \(F\) and \(H\), respectively. Hence,
\eqref{eq:rcpp_constrained_reformulation} can be written compactly as
\begin{equation}
\begin{aligned}
\min_{\mathbf x,\mathbf z}\quad
& F(\mathbf x)+H(\mathbf z)\\
\mathrm{s.t.}\quad
& \mathbf A\mathbf x-\mathbf B\mathbf z=\mathbf 0 .
\end{aligned}
\tag{P4}
\label{prob:original-constrained-problem}
\end{equation}

This formulation preserves the information theoretic structure of the RCPP problem while revealing a linearly constrained two block composite structure. The \(\mathbf x\)-block represents the marginal distribution of the released representation, with its simplex constraint included through \(F^{1}\). The \(\mathbf z\)-block represents the privacy mechanism and contains the conditional entropy terms, with its probability constraints included through \(H^{1}\). The linear constraint \(\mathbf A\mathbf x-\mathbf B\mathbf z=\mathbf 0\) enforces marginal consistency and perfect privacy. This structure provides the basis for the ADMM algorithm developed in the next section.

\section{Proposed Approach}
\label{sec:proposed}

Although~\eqref{prob:original-constrained-problem} has a two block linearly constrained form, the direct use of a standard ADMM scheme does not by itself yield the convergence analysis needed for the present problem. Existing convergence results for linearly constrained composite information theoretic problems typically rely on structural conditions such as range compatibility between the constraint matrices and smoothness of the objective term associated with the last updated primal block. These conditions are not directly satisfied by \eqref{prob:original-constrained-problem}: the variables themselves are constrained to probability sets: \(\mathbf x\) lies in the simplex \(\Delta_{\mathcal U}\), while \(\mathbf z\) lies in the set \(\mathcal Q\) of conditional probability distributions. When these constraints are incorporated into the objective through the indicator functions
\(\iota_{\Delta_{\mathcal U}}\) and \(\iota_{\mathcal Q}\), the corresponding block objectives become nonsmooth. Consequently, the block optimality conditions contain normal cone terms associated with these probability sets, and these terms have to be retained in the convergence analysis. 

Motivated by the perturbed dual-update schemes in~\cite{yang2022proximal,zhou2025perturbed}, which were developed for broad classes of nonconvex and nonsmooth optimization problems, we specialize their perturbation principle to the structure of~\eqref{prob:original-constrained-problem}, rather than seeking to extend the general problem classes considered therein. This specialization leads to a different design of the primal updates and requires a problem-specific convergence analysis. On the theoretical side as well, while~\cite{zhou2025perturbed} establishes a complexity bound by showing the existence of an approximately stationary point within the first \(K\) iterations, we go beyond the result in~\cite{zhou2025perturbed} and provide a convergence proof of the entire sequence. For instance, we exploit the K{\L} property which we prove in the paper, and derive convergence rates for both the Lyapunov error and the distance to the limit point in terms of the K{\L} exponent. Our results show that while the convergence rate is sublinear in general, linear convergence rate can be achieved under some assumptions.

In particular, the objective exhibits a strongly convex--weakly convex block structure~\cite{huang2021provably}. For the $\mathbf{x}$-block, the strong convexity of the corresponding objective component, together with the augmented quadratic term, provides sufficient curvature for the descent estimate. Hence, no additional proximal regularization is required in the $\mathbf{x}$-update. By contrast, the $\mathbf{z}$-block is generally nonconvex and nonsmooth, and it does not contain a strongly convex component that uniformly dominates its possible negative curvature. A proximal term is therefore introduced only in the $\mathbf{z}$-subproblem to provide the additional curvature and control of successive iterates required by the convergence analysis.

The smooth terms are retained in their original form rather than linearized in the ADMM subproblems. This is particularly important for the $\mathbf{x}$-block, since retaining its strongly convex component allows the associated curvature to be used directly in the descent analysis. Linearizing this component at the current outer iterate would remove that curvature from the subproblem, and an additional proximal term would generally be required to recover it, together with an extra algorithmic parameter. Although the resulting subproblems do not necessarily admit closed-form solutions, they can be approximated efficiently by projected gradient iterations over the corresponding probability sets. The implementation details are provided in Section~\ref{sec:numerical_results}.

Therefore, we introduce the following perturbed augmented Lagrangian as the basis of the proposed method:
\begin{equation}
\begin{aligned}
\mathcal L_{\rho,\tau}(\mathbf x,\mathbf z,\boldsymbol\lambda)
&\triangleq
F^{0}(\mathbf x)+F^{1}(\mathbf x)
+H^{0}(\mathbf z)+H^{1}(\mathbf z)\nonumber  \\
&\quad
+\left\langle (1-\tau)\boldsymbol\lambda,\mathbf A\mathbf x-\mathbf B\mathbf z\right\rangle
+\frac{\rho}{2}\left\|\mathbf A\mathbf x-\mathbf B\mathbf z\right\|^{2},
\end{aligned}
\label{eq:perturbed_augmented_lagrangian}
\end{equation}
where \(\rho>0\) is the penalty parameter and \(\tau\in(0,1)\) is a perturbation parameter. In addition to the perturbed multiplier update, the \(\mathbf z\)-subproblem is regularized by the proximal term
\[
\frac{\gamma}{2}\|\mathbf z-\mathbf z^k\|_{\mathbf Q}^{2},
\quad
\gamma\geq 0,\quad \mathbf Q\succeq \mathbf 0.
\]
This term controls successive changes in the probability variables and provides the additional regularization required for the subsequent descent analysis. The resulting solver is then given by:

\begin{subequations}
\label{eq:admm_updates}
\begin{align}
\mathbf x^{k+1}
&\begin{aligned}[t]
&\in \arg\min_{\mathbf x}
\Big\{
F^0(\mathbf x)+F^1(\mathbf x)
+\left\langle (1-\tau)\boldsymbol\lambda^k,\right.\\
&\quad\quad\left.\mathbf A\mathbf x-\mathbf B\mathbf z^k \right\rangle
+\frac{\rho}{2}\left\|\mathbf A\mathbf x-\mathbf B\mathbf z^k\right\|^2
\Big\},
\end{aligned}
\label{eq:x-update}\\
\mathbf z^{k+1}
&\begin{aligned}[t]
&\in \arg\min_{\mathbf z}
\Big\{
H^0(\mathbf z)+H^1(\mathbf z)
+\left\langle (1-\tau)\boldsymbol\lambda^k,\right.\\
&\quad\quad\left.\mathbf A\mathbf x^{k+1}-\mathbf B\mathbf z \right\rangle
+\frac{\rho}{2}\left\|\mathbf A\mathbf x^{k+1}-\mathbf B\mathbf z\right\|^2\\
&\quad\quad
+\frac{\gamma}{2}\|\mathbf z-\mathbf z^{k}\|_{\mathbf Q}^2
\Big\},
\end{aligned}
\label{eq:z-update}\\
\boldsymbol\lambda^{k+1}
&=(1-\tau)\boldsymbol\lambda^k
+\rho\left(\mathbf A\mathbf x^{k+1}-\mathbf B\mathbf z^{k+1}\right).
\label{eq:dual-update}
\end{align}
\end{subequations}

These updates constitute the proposed proximal perturbed ADMM method for solving~\eqref{prob:original-constrained-problem}. The complete procedure, including the initialization and termination criterion, is summarized in Algorithm~\ref{alg:admm}.
\begin{algorithm}
\caption{Proximal Perturbed ADMM Method for~\eqref{prob:original-constrained-problem}}
\label{alg:admm}
\begin{algorithmic}[1]
\REQUIRE Algorithmic parameters \(\rho\), \(\tau\), \(\gamma\) and \(\mathbf Q\) satisfying~\eqref{eq:sufficient-descent-parameter-condition}, tolerance \(\epsilon>0\), maximum iteration number \(K\).
\STATE \textbf{Initialize}
    $\mathbf x^{0}\in\Delta_{\mathcal U}$,
    $\mathbf z^{0}\in\mathcal Q$,
    $\boldsymbol\lambda^{0}$, and set $k\gets0$.
\REPEAT
    \STATE
    \[
    \mathbf x^{k+1}
    \begin{aligned}[t]
    &\in \arg\min_{\mathbf x}
    \Big\{
    F^0(\mathbf x)+F^1(\mathbf x)
    +\left\langle (1-\tau)\boldsymbol\lambda^k,\right.\\
    &\quad\quad\left.\mathbf A\mathbf x-\mathbf B\mathbf z^k \right\rangle
    +\frac{\rho}{2}\left\|\mathbf A\mathbf x-\mathbf B\mathbf z^k\right\|^2
    \Big\}.
    \end{aligned}
    \]
    \STATE
    \[
    \mathbf z^{k+1}
    \begin{aligned}[t]
    &\in \arg\min_{\mathbf z}
    \Big\{
    H^0(\mathbf z)+H^1(\mathbf z)
    +\left\langle (1-\tau)\boldsymbol\lambda^k,\right.\\
    &\quad\quad\left.\mathbf A\mathbf x^{k+1}-\mathbf B\mathbf z \right\rangle
    +\frac{\rho}{2}\left\|\mathbf A\mathbf x^{k+1}-\mathbf B\mathbf z\right\|^2\\
    &\quad\quad
    +\frac{\gamma}{2}\|\mathbf z-\mathbf z^{k}\|_{\mathbf Q}^2
    \Big\}.
    \end{aligned}
    \]
    \STATE
    \[
    \bm\lambda^{k+1}
    =
    (1-\tau)\bm\lambda^{k}
    +\rho
    \left(
    \mathbf A\mathbf x^{k+1}
    -\mathbf B\mathbf z^{k+1}
    \right).
    \]
    \STATE $k\gets k+1$.
\UNTIL{
    $k=K$ or
    $\|\mathbf A\mathbf x^{k}-\mathbf B\mathbf z^{k}\|\le\epsilon$
}
\STATE
    \textbf{return}
    $\mathbf x^{k}$, $\mathbf z^{k}$, and $\boldsymbol\lambda^{k}$.
\end{algorithmic}
\end{algorithm}

\section{Convergence Analysis}
\label{sec:convergence_analysis}

We now establish the convergence results of Algorithm~\ref{alg:admm}. For notational convenience, define
\(
\mathbf{u}^k
:=
\bigl(\mathbf x^k,\mathbf z^k,\boldsymbol\lambda^k\bigr),
\mathbf{u}^\ast
:=
\bigl(\mathbf x^\ast,\mathbf z^\ast,\boldsymbol\lambda^\ast\bigr),
\)
where \(\mathbf{u}^\ast\) denotes an accumulation point of the sequence \(\{\mathbf{u}^k\}_{k\in\mathbb N}\).
The following assumptions are imposed throughout this section.

\begin{assumption}[Feasibility]
\label{ass:feasibility}
Problem~\eqref{prob:original-constrained-problem} is feasible, and its set of stationary points is nonempty.
\end{assumption}

\begin{assumption}[Uniform positivity]
\label{ass:uniform_positivity}
There exists a constant \(\epsilon>0\) such that, for every iterate generated by Algorithm~\ref{alg:admm},
\[
p_U^k(u)\ge \epsilon,\qquad
p_{U|X}^k(u|x)\ge \epsilon,
\]
for all \(u\in\mathcal U\) and all \(x\in\mathcal X\).
\end{assumption}
Under Assumption~\ref{ass:uniform_positivity}, \(F^0\) is \(\mu_{F^0}\)-strongly convex, where \(\mu_{F^0}>0\) denotes the corresponding strong convexity modulus~\cite{huang2021provably}. Moreover, by~\cite[Lemma~3]{huang2022linearly}, \(H^0\) is
\(\omega_{H^0}\)-restricted weakly convex with respect to
\(
\mathbf B_1
:=
\mathbf p_X^{\mathsf T}\otimes\mathbf I_{|\mathcal U|},
\)
where \(\omega_{H^0}\geq0\) denotes the corresponding modulus. In our formulation, the matrix \(\mathbf B\) is given by the vertical concatenation
\[
\mathbf B
=
\begin{bmatrix}
\mathbf B_1\\
\mathbf B_2
\end{bmatrix},
\qquad
\mathbf B_2
:=
\left(
\mathbf P_{X|S}^{\mathsf T}
-
\mathbf 1_{|\mathcal S|}\mathbf p_X^{\mathsf T}
\right)
\otimes\mathbf I_{|\mathcal U|}.
\]
Since
\(
\|\mathbf B\mathbf z\|^2
=
\|\mathbf B_1\mathbf z\|^2
+
\|\mathbf B_2\mathbf z\|^2,
\)
we have
\[
H^0(\mathbf z)
+\frac{\omega_{H^0}}{2}\|\mathbf B\mathbf z\|^2
=
H^0(\mathbf z)
+\frac{\omega_{H^0}}{2}\|\mathbf B_1\mathbf z\|^2
+\frac{\omega_{H^0}}{2}\|\mathbf B_2\mathbf z\|^2.
\]
The first two terms on the right-hand side define a convex function by the restricted weak convexity of \(H^0\) with respect to \(\mathbf B_1\), while the last term is convex. Therefore, \(H^0\) is also \(\omega_{H^0}\)-restricted weakly convex with respect to the matrix \(\mathbf B\).

To establish convergence, we need a Lyapunov function that satisfies a sufficient descent property and is bounded from below. In the present setting, the augmented Lagrangian itself cannot directly serve as a Lyapunov function, since the positive term induced by the dual update cannot be directly controlled by the successive primal changes. We therefore first derive two auxiliary results to identify and control these terms.
\begin{proposition}
\label{prop:combined_descent_rearranged}
For the sequence \(\{\mathbf{u}^k\}_{k\in\mathbb N}\) generated by Algorithm~\ref{alg:admm}, we have
\begin{equation}
\begin{aligned}
&\frac{\rho}{2\eta_1}
\|\mathbf x^{k+1}-\mathbf x^k\|_{\mathbf A^\top\mathbf A}^2
-\mu_{F^0}\|\mathbf x^{k+1}-\mathbf x^k\|^2\\
&
+\|\mathbf z^{k+1}-\mathbf z^k\|_{\mathbf E_\mathbf z}^2
-\frac{\tau}{\rho}
\|\boldsymbol\lambda^{k+1}-\boldsymbol\lambda^k\|^2 \\
\ge&
\|\mathbf z^{k+1}-\mathbf z^k\|_{\mathbf D_\mathbf z}^2
-\|\mathbf z^k-\mathbf z^{k-1}\|_{\mathbf D_\mathbf z}^2 \\
&
+\frac{1-\tau}{2\rho}
\|\boldsymbol\lambda^{k+1}-\boldsymbol\lambda^k\|^2
-\frac{1-\tau}{2\rho}
\|\boldsymbol\lambda^k-\boldsymbol\lambda^{k-1}\|^2 .
\end{aligned}
\label{eq:combined-descent-rearranged}
\end{equation}
where
\[
\mathbf E_\mathbf z:=
\left(
\omega_{H^0}
+2\rho\eta_1
\right)\mathbf B^\top\mathbf B,
\quad
\mathbf D_\mathbf z:=
\rho\eta_1
\mathbf B^\top\mathbf B
+\frac{\gamma}{2}\mathbf Q .
\]
\end{proposition}
This proposition establishes an inter-iteration inequality that produces a descent term involving the current dual change, while controlling the remaining quantities by changes from the preceding iteration. This dual descent term is crucial for compensating for the positive contribution of the dual update and for constructing the Lyapunov function. The proof is provided in Appendix~\ref{app:proof_combined_descent_rearranged}.

\begin{proposition}
\label{prop:one_step_descent_AL}
For the sequence \(\{\mathbf{u}^k\}_{k\in\mathbb N}\)
generated by Algorithm~\ref{alg:admm}, we have
\begin{equation} 
    \begin{aligned} 
        &\left[ \mathcal L_{\rho,\tau} (\mathbf x^{k+1},\mathbf z^{k+1},\boldsymbol\lambda^{k+1}) -\frac{\tau(1-\tau)}{2\rho} \|\boldsymbol\lambda^{k+1}\|^2 \right] \\ 
        &\quad 
        -\left[ \mathcal L_{\rho,\tau} (\mathbf x^{k},\mathbf z^{k},\boldsymbol\lambda^{k}) -\frac{\tau(1-\tau)}{2\rho} \|\boldsymbol\lambda^{k}\|^2 \right] \\ 
        &\le -\frac{1}{2}\|\mathbf x^{k+1}-\mathbf x^k\|^2_{\mu_{F^0}\mathbf I+\rho\mathbf A^\top\mathbf A} \\ 
        &\quad -\frac{1}{2} \|\mathbf z^{k+1}-\mathbf z^k\|^2_ {(\rho-\omega_{H^0})\mathbf B^{\top}\mathbf B+2\gamma\mathbf Q} \\ 
        &\quad 
        +\frac{(1-\tau)(2-\tau)}{2\rho} \|\boldsymbol\lambda^{k+1}-\boldsymbol\lambda^k\|^2 . 
    \end{aligned} 
    \label{eq:one_step_descent_AL} 
\end{equation} 
\end{proposition}
This proposition characterizes the one-step variation of the augmented Lagrangian and reveals the positive contribution induced by the dual update. Combined with the dual descent term established in Proposition~\ref{prop:combined_descent_rearranged}, this inequality provides the basis for introducing the correction terms in the Lyapunov function defined below.
The proof is given in Appendix~\ref{app:proof_one_step_descent_AL}.

To simplify the notation used throughout the following analysis, we define
\(
\mathbf w^{k+1}:=
(\mathbf x^{k+1},\mathbf z^{k+1},\boldsymbol\lambda^{k+1},\mathbf z^k,\boldsymbol\lambda^k).
\)
Accordingly, the associated Lyapunov function \(\mathcal{P}(\mathbf{w}^{k+1})\) is defined as follows.
\begin{equation}
\begin{aligned}
&\mathcal P(\mathbf w^{k+1})
:=
\mathcal L_{\rho,\tau}(\mathbf x^{k+1},\mathbf z^{k+1},\boldsymbol\lambda^{k+1})-\frac{\tau(1-\tau)}{2\rho}\|\boldsymbol\lambda^{k+1}\|^2 \\
&+d\Bigg[
\|\mathbf z^{k+1}-\mathbf z^k\|_{\mathbf D_\mathbf z}^2+\frac{1-\tau}{2\rho}\|\boldsymbol\lambda^{k+1}-\boldsymbol\lambda^k\|^2 
\Bigg].
\end{aligned}
\label{eq:Lyapunov-function}
\end{equation}
where \(d>0\) is chosen to establish sufficient descent and lower boundedness of the Lyapunov sequence.
The correction terms in \eqref{eq:Lyapunov-function} are chosen so as to absorb the history-dependent quantities appearing in Proposition~\ref{prop:combined_descent_rearranged}. We next establish the sufficient descent and lower boundedness properties of the resulting Lyapunov sequence \(\{\mathcal P(\mathbf w^k)\}_{k\in\mathbb N}\).

\begin{lemma}[Sufficient descent of the Lyapunov sequence]
\label{lem:sufficient_descent_lyapunov}
Suppose that Assumptions~\ref{ass:feasibility}--\ref{ass:uniform_positivity} hold.
If the algorithmic parameters \(\rho\), \(\tau\), \(\gamma\), and \(\mathbf Q\) are chosen such that there exist auxiliary constants \(d>0\) and \(\eta_1>0\) satisfying
\begin{equation}
\begin{aligned}
&\rho>0,\qquad
0<\tau<1,\qquad
\gamma>0,\qquad
\mathbf Q\succ\mathbf 0,\\
&\eta_1\geq d,\qquad
d>\frac{(1-\tau)(2-\tau)}{2\tau},\\
&2\gamma\lambda_{\min}(\mathbf Q)
>
\left[
\omega_{H^0}
+2d\left(
\omega_{H^0}+2\rho\eta_1
\right)
-\rho
\right]_{+}
\|\mathbf B\|^2.
\end{aligned}
\label{eq:sufficient-descent-parameter-condition}
\end{equation}
Here, \(d\) and \(\eta_1\) are auxiliary parameters used only in the
convergence analysis.
Then, for every \(k\in\mathbb N\),
\begin{equation}
\begin{aligned}
&\mathcal P(\mathbf w^{k+1})-\mathcal P(\mathbf w^k)
\le \nonumber
-\frac{1}{2}\|\mathbf x^{k+1}-\mathbf x^k\|_{\mathbf M_\mathbf x}^2\\
&\quad
-\frac{1}{2}\|\mathbf z^{k+1}-\mathbf z^k\|_{\mathbf M_\mathbf z}^2  
-a_{\boldsymbol\lambda}\|\boldsymbol\lambda^{k+1}-\boldsymbol\lambda^k\|^2 ,
\end{aligned}
\label{eq:sufficient-descent-lemma}
\end{equation}
where
\[
\begin{aligned}
&\mathbf E_{\mathbf z}:=
\left(
\omega_{H^0}
+2\rho\eta_1
\right)\mathbf B^\top\mathbf B,\\
&\mathbf M_{\mathbf x}
:=
\left(1+2d\right)\mu_{F^0}\mathbf I+\left(1-\frac{d}{\eta_1}\right)\rho\mathbf A^\top\mathbf A\succ0,\\
&\mathbf M_{\mathbf z}:=
(\rho-\omega_{H^0})\mathbf B^\top\mathbf B
+2\gamma\mathbf Q
-2d\mathbf E_{\mathbf z}\succ0,\\
&a_{\boldsymbol\lambda}:=
\frac{d\tau}{\rho}
-\frac{(1-\tau)(2-\tau)}{2\rho}>0.
\end{aligned}
\]
Consequently, the Lyapunov sequence \(\{\mathcal P(\mathbf w^k)\}_{k\in\mathbb N}\) is monotonically nonincreasing, and there exists a constant \(C_1>0\) such that
\begin{equation}
\begin{aligned}
&\mathcal P(\mathbf w^{k+1})-\mathcal P(\mathbf w^k)\\
&\le
-C_1\left(
\|\mathbf x^{k+1}-\mathbf x^k\|^2
+\|\mathbf z^{k+1}-\mathbf z^k\|^2
+\|\boldsymbol\lambda^{k+1}-\boldsymbol\lambda^k\|^2
\right),
\label{eq:sufficient-descent-C1}
\end{aligned}
\end{equation}
where
\[
C_1:=
\min\left\{
\frac{1}{2}\lambda_{\min}(\mathbf M_{\mathbf x}),
\frac{1}{2}\lambda_{\min}(\mathbf M_{\mathbf z}),
a_{\boldsymbol\lambda}
\right\}>0.
\]
\end{lemma}
The proof is given in Appendix~\ref{app:proof_sufficient_descent_lyapunov}.

\begin{lemma}[Lower boundedness of the Lyapunov sequence]
\label{lem:lower_boundedness_lyapunov}
Suppose that Assumptions~\ref{ass:feasibility}--\ref{ass:uniform_positivity} hold and that the parameter conditions in Lemma~\ref{lem:sufficient_descent_lyapunov} are satisfied. Then the dual sequence \(\{\|\boldsymbol\lambda^k\|\}_{k\in\mathbb N}\) is bounded from above. Moreover, the Lyapunov sequence \(\{\mathcal P(\mathbf w^k)\}_{k\in\mathbb N}\) is bounded from below.
\end{lemma}
The proof is given in Appendix~\ref{app:proof_lower_boundedness_lyapunov}.

The above lemmas establish the two key properties of the Lyapunov sequence required for the subsequent convergence analysis. Since the perturbed dual update may leave a nonzero asymptotic feasibility residual when \(\tau>0\), we first introduce the following \(\epsilon\)-KKT condition to characterize the accumulation points of the generated sequence.
\begin{definition}
\label{def:epsilon_akkt_kkt}
Consider the following inequalities with $\epsilon\geq0$:
\begin{subequations}
\begin{align}
\operatorname{dist}\!\left(\mathbf 0,\nabla F^0(\mathbf x^\ast)+\partial F^1(\mathbf x^\ast)+\mathbf A^\top\boldsymbol\lambda^\ast\right)
&\le \epsilon, \label{eq:epsilon-kkt-x}\\
\operatorname{dist}\!\left(\mathbf 0,\nabla H^0(\mathbf z^\ast)+\partial H^1(\mathbf z^\ast)-\mathbf B^\top\boldsymbol\lambda^\ast\right)
&\le \epsilon, \label{eq:epsilon-kkt-z}\\
\left\|\mathbf A\mathbf x^\ast-\mathbf B\mathbf z^\ast\right\|
&\le \epsilon. \label{eq:epsilon-kkt-feas}
\end{align}
\end{subequations}
If a point $(\mathbf x^\ast,\mathbf z^\ast,\boldsymbol\lambda^\ast)$ satisfies
\eqref{eq:epsilon-kkt-x}, \eqref{eq:epsilon-kkt-z}, and
\eqref{eq:epsilon-kkt-feas}, it is referred to as an $\epsilon$-KKT point.
\end{definition}

The following theorem establishes boundedness and asymptotic regularity of the generated sequence \(\{\mathbf u^k\}_{k\in\mathbb N}\) and characterizes each of its accumulation points in terms of the above approximate stationarity condition.
\begin{theorem}[Subsequential convergence and approximate stationarity]
\label{thm:subsequential_convergence}
Suppose that Assumptions~\ref{ass:feasibility}--\ref{ass:uniform_positivity} hold and the parameter conditions in Lemma~\ref{lem:sufficient_descent_lyapunov} are satisfied. Then the sequence \(\{\mathbf u^k\}_{k\in\mathbb N}\) is bounded and therefore has at least one accumulation point. Moreover, the successive differences vanish, namely,
\[
\mathbf x^{k+1}-\mathbf x^k\to0,\quad
\mathbf z^{k+1}-\mathbf z^k\to0,\quad
\boldsymbol\lambda^{k+1}-\boldsymbol\lambda^k\to0 .
\]
Furthermore, the asymptotic feasibility residual satisfies
\[
\limsup_{k\to\infty}
\|\mathbf A\mathbf x^{k+1}-\mathbf B\mathbf z^{k+1}\|
\le
\frac{\tau}{\rho}
\limsup_{k\to\infty}\|\boldsymbol\lambda^k\|.
\]
Consequently, every accumulation point \(\mathbf u^\ast\) is an \(\epsilon\)-KKT point of~\eqref{prob:original-constrained-problem}, with
\[
\epsilon:=
\frac{\tau}{\rho}
\limsup_{k\to\infty}\|\boldsymbol\lambda^k\|.
\]
\end{theorem}
The proof is given in Appendix~\ref{app:proof_subsequential_convergence}.

\section{Convergence Rate Analysis}
\label{sec:convergence_rate_analysis}
The previous section established boundedness, subsequential convergence, and approximate stationarity of the sequence generated by Algorithm~\ref{alg:admm}. In this section, we strengthen these results under the Kurdyka-{\L}ojasiewicz framework by proving convergence of the entire sequence and deriving its asymptotic rates. 

\begin{lemma}[Subgradient bound]
\label{lem:subgradient_bound}
Suppose that Assumptions~\ref{ass:feasibility}--\ref{ass:uniform_positivity} hold and the parameter conditions in Lemma~\ref{lem:sufficient_descent_lyapunov} are satisfied. Then there exists a constant \(C_2>0\) such that, for every \(k\in\mathbb N\),
\[
\begin{aligned}
&\operatorname{dist}\!\left(\mathbf 0,\partial_\mathbf w\mathcal P(\mathbf w^{k+1})\right)\\
&\le\mspace{-2 mu}
C_2 \mspace{-2 mu}
\left(
\|\mathbf x^{k+1}-\mathbf x^k\|^2
+\|\mathbf z^{k+1}-\mathbf z^k\|^2
+\|\boldsymbol\lambda^{k+1}-\boldsymbol\lambda^k\|^2
\right)^{1/2}.
\end{aligned}
\]
\end{lemma}
The proof is given in Appendix~\ref{app:proof_subgradient_bound}.

\begin{lemma}[K{\L} property of \(\mathcal P\)]
\label{lem:KL_property}
Suppose that Assumptions~\ref{ass:feasibility}--\ref{ass:uniform_positivity} hold, then Lyapunov function \(\mathcal P\) is a K{\L} function. In particular, for any critical point \(\mathbf w^\ast\in\operatorname{dom}\partial\mathcal P\), there exist constants \(C_3>0\), \(\theta\in[0,1)\), \(\nu>0\), and a neighborhood \(\mathcal N(\mathbf w^\ast)\) of \(\mathbf w^\ast\) such that
\begin{equation}
\operatorname{dist}\!\left(\mathbf 0,\partial_\mathbf w\mathcal P(\mathbf w)\right)
\ge
C_3\left(\mathcal P(\mathbf w)-\mathcal P(\mathbf w^\ast)\right)^\theta
\label{eq:KL-property-general}
\end{equation}
for all \(\mathbf w\in\mathcal N(\mathbf w^\ast)\) satisfying
\(
\mathcal P(\mathbf w^\ast)<\mathcal P(\mathbf w)<\mathcal P(\mathbf w^\ast)+\nu .
\)
\end{lemma}
The proof is given in Appendix~\ref{app:proof_KL_property}.

By Theorem~\ref{thm:subsequential_convergence}, there exists a
subsequence $\{\mathbf u^{k_j}\}_{j\in\mathbb N}$ such that
$\mathbf u^{k_j}\to\mathbf u^*$. Since the successive differences vanish, we also have
$\mathbf z^{k_j-1}\to\mathbf z^*$ and
$\boldsymbol\lambda^{k_j-1}\to\boldsymbol\lambda^*$.
Thus,
\(
\mathbf w^{k_j}\to\mathbf w^*.
\)
The indicator terms in $\mathcal P$ vanish along this subsequence and at its limit, and all remaining terms are continuous. Hence,
\(
\mathcal P(\mathbf w^{k_j})\to\mathcal P(\mathbf w^*),
\)
which verifies the limiting continuity condition in~\cite{Attouch2013}. 
Lemmas~\ref{lem:sufficient_descent_lyapunov}, \ref{lem:lower_boundedness_lyapunov}, \ref{lem:KL_property}, together with the limiting continuity yield the following result. 
\begin{lemma}[Finite length property and the whole sequence convergence{\normalfont\protect~\cite[Theorem~2.9]{Attouch2013}}]
\label{lemma:finite-length}
Suppose that Assumptions~\ref{ass:feasibility}--\ref{ass:uniform_positivity} hold and that the parameter conditions in Lemma~\ref{lem:sufficient_descent_lyapunov} are satisfied. Then the sequence $\{\mathbf w^k\}_{k\in\mathbb N}$ has a finite length, i.e.,
\[
\sum_{k=0}^{+\infty}
\|\mathbf w^{k+1}-\mathbf w^k\|<+\infty,
\]
and the generated sequence converges to an unique accumulation point
$\mathbf w^*$.
\end{lemma}

Following the standard K{\L} analysis framework in~\cite{Attouch2013,Frankel2015}, we combine the sufficient descent inequality, the subgradient bound, and the K{\L} inequality to derive a recursion for the Lyapunov error. Let \(e_k:=\mathcal P(\mathbf w^k)-\mathcal P(\mathbf w^\ast).\) The following lemma gives the resulting recursion.
\begin{lemma}
\label{lem:descent_subgradient_recursion}
Suppose that Assumptions~\ref{ass:feasibility}--\ref{ass:uniform_positivity} hold and the parameter conditions in Lemma~\ref{lem:sufficient_descent_lyapunov} are satisfied. Then there exist \(k_0\in\mathbb N\), \(\theta\in[0,1)\), and \(\bar C:=\frac{C_1C_3^2}{C_2^2}>0\) such that
\begin{equation}
e_k-e_{k+1}
\ge
\bar C e_{k+1}^{2\theta},
\quad
\forall \, k\in\mathbb N,\ k\ge k_0.
\label{eq:KL-recursion-lemma}
\end{equation}
\end{lemma}

The proof is given in Appendix~\ref{app:proof_descent_subgradient_recursion}.

We now use the above relation to establish the convergence rates of the entire generated sequence, and the Lyapunov error according to the K{\L} exponent.

\begin{theorem}[Convergence rate of the Lyapunov error sequence \(\{e_k\}_{k\in\mathbb N}\)]
\label{thm:rate_lyapunov_error}
Suppose that Assumptions~\ref{ass:feasibility}--\ref{ass:uniform_positivity} hold and that the parameter conditions in Lemma~\ref{lem:sufficient_descent_lyapunov} are satisfied, let $\mathbf w^\ast$ be the limit established in Lemma~\ref{lemma:finite-length}. Then the Lyapunov error sequence \(\{e_k\}_{k\in\mathbb N}\)
satisfies the following rates:
\begin{enumerate}
    \item[(i)] If \(\theta=0\), then \(e_k=0\) for all sufficiently large \(k\).

    \item[(ii)] If \(\theta\in(0,1/2]\), then
    \begin{equation}
    e_k
    \le
    e_{k_0}
    \left(1+\bar C e_{k_0}^{2\theta-1}\right)^{-(k-k_0)},~\forall \, k\in\mathbb N,\ k\ge k_0.
    \label{eq:linear-rate-ek-explicit}
    \end{equation}

    \item[(iii)] If \(\theta\in(1/2,1)\), then there exists a constant \(\mu>0\) such that
    \begin{equation}
    e_k
    \le
    \left(\mu(k-k_0)+e_{k_0}^{1-2\theta}\right)^{-\frac{1}{2\theta-1}},~\forall \, k\in\mathbb N,\ k\ge k_0 .
    \label{eq:sublinear-rate-ek-theorem}
    \end{equation}
\end{enumerate}
\end{theorem}
The proof is given in Appendix~\ref{app:proof_rate_lyapunov_error}.

\begin{theorem}[Convergence rate of the sequence \(\{\mathbf u^k\}_{k\in\mathbb N}\)]
\label{thm:rate_iterates}
Suppose that Assumptions~\ref{ass:feasibility}--\ref{ass:uniform_positivity} hold and that the parameter conditions in Lemma~\ref{lem:sufficient_descent_lyapunov} are satisfied, there exists a constant \(C>0\) such that the following rates hold:
\begin{enumerate}
    \item[(i)] If \(\theta=0\), then \(\mathbf u^k=\mathbf u^\ast\) for all sufficiently large \(k\).

    \item[(ii)] If \(\theta\in(0,1/2]\), then \(\forall k \in \mathbb N,\ k\ge k_0+1\),
    \begin{equation}
    \|\mathbf u^k-\mathbf u^\ast\|
    \le
    C\sqrt{e_{k_0}}
    \left(
    1+\bar C e_{k_0}^{2\theta-1}
    \right)^{-\frac{k-1-k_0}{2}}.
    \label{eq:linear-rate-uk-theorem}
    \end{equation}

    \item[(iii)] If \(\theta\in(1/2,1)\), then \(\forall k \in \mathbb N,\ k\ge k_0+1\),
    \begin{equation}
    \|\mathbf u^k-\mathbf u^\ast\|
    \le
    C
    \left(
    \mu(k-1-k_0)+e_{k_0}^{1-2\theta}
    \right)^{-\frac{1-\theta}{2\theta-1}}.
    \label{eq:sublinear-rate-uk-theorem}
    \end{equation}
\end{enumerate}
\end{theorem}
The proof is given in Appendix~\ref{app:proof_rate_iterates}.

The preceding theorem characterizes the local convergence rate of the generated sequence \(\{\mathbf u^k\}_{k\in\mathbb N}\) toward their limit. We next use this estimate to obtain a finite iteration threshold beyond which every subsequent iterate satisfies a prescribed approximate KKT condition. Since \(\theta\in(1/2,1)\) corresponds to the sublinear regime and provides the weakest convergence guarantee among the three cases, we state the result for this regime.
\begin{corollary}[Eventual \(\epsilon\)-KKT guarantee]
\label{cor:eventual-eKKT}
Suppose that Assumptions~\ref{ass:feasibility}--\ref{ass:uniform_positivity} hold and the parameter conditions in Lemma~\ref{lem:sufficient_descent_lyapunov} are satisfied. With \(\theta\in(1/2,1)\).
Then there exist constants \(C_{\mathrm{KKT}}>0\) and
\(k_0\in\mathbb N\) such that, for any prescribed accuracy
\(\bar\epsilon>\epsilon\), every \(\mathbf u^{k+1}\), \(k\geq K_{\bar\epsilon}\ , k\in\mathbb N\), is a \(\bar\epsilon\)-KKT point with
\[
K_{\bar\epsilon}
:=
k_0+1+
\left\lceil
\frac{1}{\mu}
\left[
\left(
\frac{C_{\mathrm{KKT}}}
{\bar\epsilon-\epsilon}
\right)^{\frac{2\theta-1}{1-\theta}}
-
e_{k_0}^{1-2\theta}
\right]_{+}
\right\rceil.
\]
\end{corollary}
The proof is provided in Appendix~\ref{app:proof_eventual-eKKT}.
\section{Inexact Subproblem Error Analysis}
\label{sec:inexact_subproblem_analysis}

We next consider the practical case where the two primal subproblems are solved only approximately. Since the preceding convergence analysis relies on the first-order optimality conditions of the subproblems, rather than on their exact global solutions, the analysis can be extended by allowing controlled residuals in these conditions.

For each \(k\in\mathbb N\), let \(\mathbf e_\mathbf x^{k+1}\) and \(\mathbf e_\mathbf z^{k+1}\) denote the residuals associated with the \(\mathbf x\)- and \(\mathbf z\)-subproblems, respectively. The inexact updates are assumed to satisfy
\begin{subequations}
\begin{align}
\mathbf 0
&\in
\nabla F^{0}(\mathbf x^{k+1})
+\partial F^{1}(\mathbf x^{k+1})
+(1-\tau)\mathbf A^{\top}\boldsymbol\lambda^{k} \notag\\
&\quad
+\rho\mathbf A^{\top}
\left(\mathbf A\mathbf x^{k+1}-\mathbf B\mathbf z^{k}\right)
+\mathbf e_\mathbf x^{k+1},
\\
\mathbf 0
&\in
\nabla H^{0}(\mathbf z^{k+1})
+\partial H^{1}(\mathbf z^{k+1})
-(1-\tau)\mathbf B^{\top}\boldsymbol\lambda^{k} \notag\\
&\quad
-\rho\mathbf B^{\top}
\left(\mathbf A\mathbf x^{k+1}-\mathbf B\mathbf z^{k+1}\right)
+\gamma\mathbf Q(\mathbf z^{k+1}-\mathbf z^{k})
+\mathbf e_\mathbf z^{k+1}.
\end{align}
\label{eq:inexact_first_order_conditions}
\end{subequations}
Moreover, the residuals are assumed to be square summable:
\begin{equation}
\sum_{k=0}^{\infty}
\left(
\|\mathbf e_\mathbf x^{k+1}\|^2
+
\|\mathbf e_\mathbf z^{k+1}\|^2
\right)
<+\infty .
\label{eq:summable-subproblem-errors}
\end{equation}
The following proposition shows that the convergence guarantee for the exact updates is preserved under this condition.
\begin{proposition}[Convergence under inexact block updates]
\label{prop:inexact-convergence}
Suppose that Assumptions~\ref{ass:feasibility}--\ref{ass:uniform_positivity} hold and the parameter conditions in Lemma~\ref{lem:sufficient_descent_lyapunov} are satisfied. If the inexact updates satisfy~\eqref{eq:inexact_first_order_conditions} and~\eqref{eq:summable-subproblem-errors}, then the conclusions of Theorem~\ref{thm:subsequential_convergence} remain valid.
\end{proposition}
The proof is given in Appendix~\ref{app:proof_inexact_sufficient_descent}.

\section{Numerical results}
\label{sec:numerical_results}
In this section, we evaluate the proposed algorithm on a discrete rate-constrained perfect-privacy information bottleneck problem. The experiments are designed to assess both the achievable rate--utility tradeoff and the effectiveness of the imposed perfect-privacy constraint. All mutual information quantities are reported in nats.

The experiments are conducted using the synthetic distribution adopted in~\cite{sreekumar2019optimal}. The corresponding joint distribution \(p_{S,X,Y}\) is given by
\[
\mathbf{P}_{S X}=0.125
\begin{bmatrix}
1&1&0&0\\
1&1&0&0\\
0&0&1&1\\
0&0&1&1
\end{bmatrix},
\quad
\mathbf P_{Y|X}=
\begin{bmatrix}
0.4&0.6\\
0.2&0.8\\
0.3&0.7\\
0.1&0.9
\end{bmatrix}^{\top}.
\]
and the representation alphabet is chosen as
\(
|\mathcal U|=2.
\)
The proposed perturbed proximal ADMM is implemented using inexact block updates. The subproblems are solved by projected gradient descent method onto probabilistic simplex~\cite{duchi2008efficient}. 

The proposed method is compared with three representative algorithms for the conventional information bottleneck problem, namely the perturbed ADMM IB algorithm, the iterative IB algorithm~\cite{tishby2000informationbottleneckmethod}, and the Douglas--Rachford splitting (DRS) IB algorithm~\cite{huang2022linearly}. The perturbed ADMM IB algorithm uses the same update scheme as the proposed method, but is applied to the conventional IB problem obtained by removing the perfect privacy constraint. These methods optimize the standard rate--utility tradeoff without explicitly enforcing the perfect-privacy constraint. For all methods, the tradeoff parameter is varied to generate the corresponding operating points on the rate--utility curve.

\begin{figure}[!t]
\centering
\includegraphics[trim=0.26cm 0.2cm 0.2cm 0.2cm, clip, width=3.49in]
{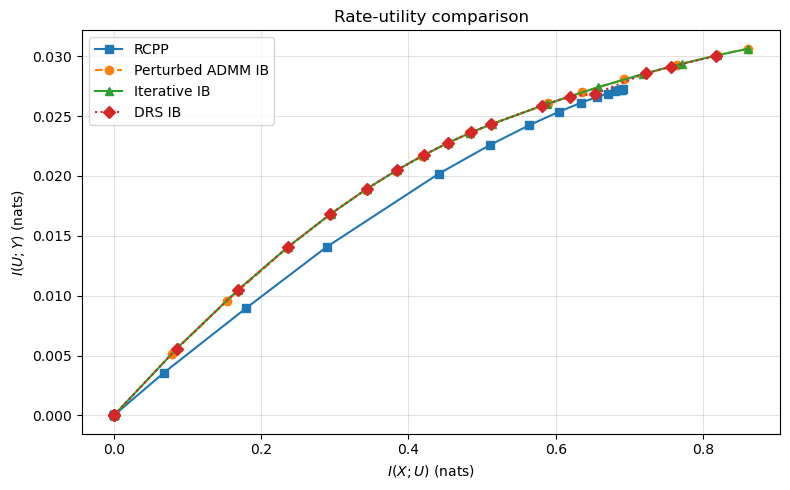}
\caption{Rate--utility performance comparison among the RCPP solution and the considered IB algorithms.}
\label{fig:rate_utility_comparison}
\label{fig_1}
\end{figure}

Fig.~\ref{fig:rate_utility_comparison} compares the rate--utility tradeoffs obtained by the proposed RCPP method and the three conventional IB algorithms. The RCPP curve is generated by solving the proposed problem under different representation-rate constraints. Thus, each point on the blue curve represents the utility achieved for a different rate limit \(R\), while satisfying the perfect-privacy constraint \(I(S;U)=0\). By varying \(R\), the proposed method traces out the rate--utility tradeoff under perfect privacy and illustrates how the representation-rate constraint affects the attainable utility.

As expected, \(I(U;Y)\) increases with \(I(X;U)\) for all the considered methods. The three conventional IB algorithms produce nearly overlapping curves over the tested range, indicating that they obtain comparable operating points for the standard IB formulation. The RCPP curve follows the same overall trend but generally lies below other curves. This difference is expected because the RCPP representation need to satisfy the additional constraint \(I(S;U)=0\).

Nevertheless, the utility gap remains relatively small over the tested rate range, indicating that perfect privacy incurs only a moderate utility loss for the considered distribution. Moreover, as the rate limit increases, the RCPP solution approaches the perfect-privacy utility reported in~\cite{sreekumar2019optimal}, indicating that the rate constraint gradually becomes inactive.

\begin{figure}[!t]
\centering
\includegraphics[width=3.45in]{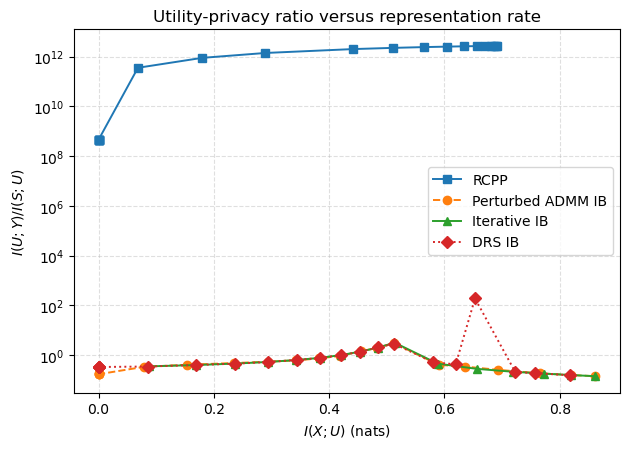}
\caption{Utility-to-privacy ratio versus the rate for the RCPP solution and the considered IB algorithms.}
\label{fig:utility_privacy_ratio}
\label{fig_2}
\end{figure}

To further illustrate the privacy-preserving capability of the proposed method, Fig.~\ref{fig:utility_privacy_ratio} reports the utility-to-privacy ratio
\[
\frac{I(U;Y)}
{\max\{I(U;S),10^{-12}\}},
\]
where the numerical floor is introduced only to avoid division by zero when the computed privacy leakage reaches machine precision. The vertical axis is shown on a logarithmic scale. Since the proposed method explicitly enforces \(I(S;U)=0\), the computed privacy leakage remains at numerical precision over the entire range of representation rates, resulting in utility-to-privacy ratios that are several orders of magnitude larger than those achieved by the conventional IB algorithms. In contrast, the baseline IB methods optimize utility without directly controlling the information leakage about the private variable, and consequently their utility-to-privacy ratios remain close to unity over most operating points. These results demonstrate that the proposed method effectively enforces perfect privacy while maintaining competitive rate--utility performance.

Overall, the numerical results validate the effectiveness of the proposed algorithm. The algorithm produces high-quality solutions that closely approach the utility achieved by conventional information bottleneck methods while simultaneously satisfying the perfect-privacy constraint to numerical precision.

\section{Conclusion}
\label{sec:conclusion}

In this work, we proposed a perturbed proximal ADMM solver for the information bottleneck problem with a rate constraint and a perfect privacy constraint over finite alphabets, with particular emphasis on the regime where the rate constraint is active. Unlike formulations based on a leakage penalty, the proposed method computes perfect privacy mechanisms by enforcing the zero leakage constraint directly while retaining an explicit information rate constraint. The resulting constrained problem is nonconvex and nonsmooth, and involves rank deficient linear constraints induced by the probability and perfect privacy requirements.

We proved convergence of the proposed algorithm through a Lyapunov analysis that accounts for the probability constraints in the constrained formulation. We further obtained convergence rate results based on the K{\L} property and extended the convergence guarantee to inexact block updates. An interesting open problem is the extension of the proposed framework to multiuser settings and the investigation of the resulting information-theoretic optimization problems.

\appendices
\section{Proof of Proposition~\ref{prop:combined_descent_rearranged}}
\label{app:proof_combined_descent_rearranged}

The first-order optimality conditions of the two primal subproblems in~\eqref{eq:x-update} and~\eqref{eq:z-update} are given by
\begin{subequations}
\label{eq:first_order_conditions}
\begin{align}
&\nabla F^{0}(\mathbf x^{k+1})+\boldsymbol\nu^{F^{1}}_{\mathbf x^{k+1}}
+(1-\tau) \mathbf A^{\top}\boldsymbol\lambda^{k} \notag\\
&\qquad
+\rho \mathbf A^{\top}\left(\mathbf A\mathbf x^{k+1}-\mathbf B\mathbf z^{k}\right)=0,
\label{eq:first_order_condition_x}\\
&\nabla H^{0}(\mathbf z^{k+1})+\boldsymbol\nu^{H^{1}}_{\mathbf z^{k+1}}
-(1-\tau) \mathbf B^{\top}\boldsymbol\lambda^{k} \notag\\
&\qquad
-\rho \mathbf B^{\top}\left(\mathbf A\mathbf x^{k+1}-\mathbf B\mathbf z^{k+1}\right)
+\gamma \mathbf Q(\mathbf z^{k+1}-\mathbf z^{k})=0.
\label{eq:first_order_condition_z}
\end{align}
\end{subequations}
where
\[
\begin{aligned}
\boldsymbol\nu^{F^{1}}_{\mathbf x^{k+1}}
&\in \partial F^{1}(\mathbf x^{k+1})
=N_{\Delta_{\mathcal U}}(\mathbf x^{k+1}),\\
\boldsymbol\nu^{H^{1}}_{\mathbf z^{k+1}}
&\in \partial H^{1}(\mathbf z^{k+1})
=N_{\mathcal Q}(\mathbf z^{k+1}).
\end{aligned}
\]
Using the dual update~\eqref{eq:dual-update}, the optimality condition of the $\mathbf x$-subproblem~\eqref{eq:x-update} can be rewritten as
\[
\begin{aligned}
&\left\langle
\nabla F^0(\mathbf x^{k+1})+\mathbf A^\top\boldsymbol\lambda^{k+1}
+\rho \mathbf A^\top \mathbf B(\mathbf z^{k+1}-\mathbf z^k),
\mathbf x-\mathbf x^{k+1}
\right\rangle \\
&=-\left\langle \boldsymbol\nu^{F^1}_{\mathbf x^{k+1}},\mathbf x-\mathbf x^{k+1}\right\rangle .
\end{aligned}
\]
Similarly, in iteration $k$, we have
\[
\begin{aligned}
&\left\langle
\nabla F^0(\mathbf x^k)+\mathbf A^\top\boldsymbol\lambda^k
+\rho \mathbf A^\top \mathbf B(\mathbf z^k-\mathbf z^{k-1}),
\mathbf x-\mathbf x^k
\right\rangle \\
&=-\left\langle \boldsymbol\nu^{F^1}_{\mathbf x^k},\mathbf x-\mathbf x^k\right\rangle .
\end{aligned}
\]
Taking $\mathbf x=\mathbf x^k$ in the first relation and $\mathbf x=\mathbf x^{k+1}$ in the second relation, and then adding them together, we obtain
\begin{equation}
\begin{aligned}
&\left\langle
\nabla F^0(\mathbf x^{k+1})-\nabla F^0(\mathbf x^k)
+\mathbf A^\top(\boldsymbol\lambda^{k+1}-\boldsymbol\lambda^k),
\mathbf x^k-\mathbf x^{k+1}
\right\rangle\\
&\quad
+\rho\left\langle
\mathbf A^\top \mathbf B\!\left[(\mathbf z^{k+1}-\mathbf z^k)-(\mathbf z^k-\mathbf z^{k-1})\right],
\mathbf x^k-\mathbf x^{k+1}
\right\rangle\\
&=
\left\langle
\boldsymbol\nu^{F^1}_{\mathbf x^k}-\boldsymbol\nu^{F^1}_{\mathbf x^{k+1}},
\mathbf x^k-\mathbf x^{k+1}
\right\rangle \ge 0,
\end{aligned}
\label{eq:x-subproblem-monotonicity}
\end{equation}
where the last inequality follows from the monotonicity of the subdifferential of the closed proper convex function $F^1$~\cite[Corollary 31.5.2]{rockafellar1997convex}.

Similarly, by the first-order optimality conditions of the \(\mathbf z\)-subproblem at iterations \(k+1\) and \(k\), and following the same derivation as above, we obtain
\begin{equation}
\begin{aligned}
&\left\langle
\nabla H^0(\mathbf z^{k+1})-\nabla H^0(\mathbf z^k)
-\mathbf B^\top(\boldsymbol\lambda^{k+1}-\boldsymbol\lambda^k),
\mathbf z^k-\mathbf z^{k+1}
\right\rangle \\
&\quad
+\left\langle
\gamma \mathbf Q\big[(\mathbf z^{k+1}-\mathbf z^k)-(\mathbf z^k-\mathbf z^{k-1})\big],
\mathbf z^k-\mathbf z^{k+1}
\right\rangle \\
&=\left\langle \boldsymbol\nu^{H^1}_{\mathbf z^k}-\boldsymbol\nu^{H^1}_{\mathbf z^{k+1}},\mathbf z^k-\mathbf z^{k+1}\right\rangle \ge 0 .
\end{aligned}
\label{eq:z-subproblem-monotonicity}
\end{equation}

Now we will analyze the LHS of~\eqref{eq:x-subproblem-monotonicity} and~\eqref{eq:z-subproblem-monotonicity}. 

Applying the strong monotonicity relation in Remark~\ref{rem:strong-convexity} to \(F^0\), with
\(\mathbf{x}_1=\mathbf{x}^{k+1}\),
\(\mathbf{x}_2=\mathbf{x}^{k}\),
\(\mathbf{u}_1=\nabla F^0(\mathbf{x}^{k+1})\), and
\(\mathbf{u}_2=\nabla F^0(\mathbf{x}^{k})\), yields
\begin{equation}
\left\langle \nabla F^0(\mathbf x^{k+1})-\nabla F^0(\mathbf x^k),\mathbf x^k-\mathbf x^{k+1}\right\rangle
\mspace{-2 mu}\le \mspace{-2 mu}-\mu_{F^0}\|\mathbf x^{k+1}-\mathbf x^k\|^2 .
\label{eq:F0-strong-convexity-bound}
\end{equation}

Moreover, for any $\eta_1>0$, Young's inequality gives
\begin{equation}
\begin{aligned}
&\rho\left\langle
\mathbf A^\top\mathbf B
\left[
(\mathbf z^{k+1}-\mathbf z^k)-(\mathbf z^k-\mathbf z^{k-1})
\right],
\mathbf x^k-\mathbf x^{k+1}
\right\rangle \\
\le&
\frac{\rho\eta_1}{2}
\left\|
\mathbf B
\left[
(\mathbf z^{k+1}-\mathbf z^k)-(\mathbf z^k-\mathbf z^{k-1})
\right]
\right\|^2\\
&
+\frac{\rho}{2\eta_1}
\left\|\mathbf A(\mathbf x^{k+1}-\mathbf x^k)\right\|^2 \\
\le&
\rho\eta_1
\|\mathbf B\mathbf z^{k+1}-\mathbf B\mathbf z^k\|^2 +\rho\eta_1
\|\mathbf B\mathbf z^k-\mathbf B\mathbf z^{k-1}\|^2\\
&
+\frac{\rho}{2\eta_1}
\|\mathbf x^{k+1}-\mathbf x^k\|_{\mathbf A^\top\mathbf A}^2 .
\end{aligned}
\label{eq:young-cross-term-xz}
\end{equation}
where the second inequality follows from
\(
\|\mathbf a-\mathbf b\|^2
\leq
2\|\mathbf a\|^2+2\|\mathbf b\|^2.
\)

Since $H^0$ is $\omega_{H^0}$-restricted weakly convex with respect to $\mathbf B$, from Remark~\ref{rem:restricted-weak-convexity}, we have
\begin{equation}
\left\langle \nabla H^0(\mathbf z^{k+1})\mspace{-2 mu}-\mspace{-2 mu}\nabla H^0(\mathbf z^k),\mathbf z^k \mspace{-4 mu}- \mspace{-4 mu}\mathbf z^{k+1}\right\rangle
\mspace{-2 mu}\le \mspace{-2 mu}\omega_{H^0}\|\mathbf B\mathbf z^{k+1}-\mathbf B\mathbf z^k\|^2 .
\label{eq:H0-restricted-weak-convexity-bound}
\end{equation}

Using the weighted norm identity
\(
2\langle \mathbf Q(\mathbf a-\mathbf b),\mathbf a\rangle
=
\|\mathbf a\|_{\mathbf Q}^2
-\|\mathbf b\|_{\mathbf Q}^2
+\|\mathbf a-\mathbf b\|_{\mathbf Q}^2,
\)
we get
\begin{equation}
\begin{aligned}
&\left\langle
\gamma \mathbf Q\big[(\mathbf z^{k+1}-\mathbf z^k)-(\mathbf z^k-\mathbf z^{k-1})\big],
\mathbf z^k-\mathbf z^{k+1}
\right\rangle \\
&= -\frac{\gamma}{2}\|\mathbf z^{k+1}-\mathbf z^k\|_{\mathbf Q}^2
+\frac{\gamma}{2}\|\mathbf z^k-\mathbf z^{k-1}\|_{\mathbf Q}^2\\
&\quad
-\frac{\gamma}{2}
\left\|(\mathbf z^{k+1}-\mathbf z^k)-(\mathbf z^k-\mathbf z^{k-1})\right\|_{\mathbf Q}^2 .
\end{aligned}
\label{eq:z-proximal-cross-term-identity}
\end{equation}

Next, we analyze the dual-related terms.
\begin{equation}
\begin{aligned}
&\left\langle \mathbf A^\top(\boldsymbol\lambda^{k+1}-\boldsymbol\lambda^k),\mathbf x^k-\mathbf x^{k+1}\right\rangle\\
&+\left\langle -\mathbf B^\top(\boldsymbol\lambda^{k+1}-\boldsymbol\lambda^k),\mathbf z^k-\mathbf z^{k+1}\right\rangle \\
&\overset{\eqref{eq:dual-update}}{=}\frac{1}{\rho}\left\langle \boldsymbol\lambda^{k+1}-\boldsymbol\lambda^k,
\boldsymbol\lambda^k-(1-\tau)\boldsymbol\lambda^{k-1}-\boldsymbol\lambda^{k+1}+(1-\tau)\boldsymbol\lambda^k
\right\rangle \\
&=-\frac{1}{\rho}\|\boldsymbol\lambda^{k+1}-\boldsymbol\lambda^k\|^2
+\frac{1-\tau}{\rho}
\left\langle \boldsymbol\lambda^{k+1}-\boldsymbol\lambda^k,\boldsymbol\lambda^k-\boldsymbol\lambda^{k-1}\right\rangle\\
&=-\frac{1+\tau}{2\rho}\|\boldsymbol\lambda^{k+1}-\boldsymbol\lambda^k\|^2
+\frac{1-\tau}{2\rho}\|\boldsymbol\lambda^k-\boldsymbol\lambda^{k-1}\|^2\\
&\quad
-\frac{1-\tau}{2\rho}
\left\|(\boldsymbol\lambda^{k+1}-\boldsymbol\lambda^k)-(\boldsymbol\lambda^k-\boldsymbol\lambda^{k-1})\right\|^2,
\end{aligned}
\label{eq:dual-cross-term-identity}
\end{equation}
where the last equality follows from the identity
\(
2\langle \mathbf a,\mathbf b\rangle=\|\mathbf a\|^2+\|\mathbf b\|^2-\|\mathbf a-\mathbf b\|^2
\)
with $\mathbf a=\boldsymbol\lambda^{k+1}-\boldsymbol\lambda^k$ and $\mathbf b=\boldsymbol\lambda^k-\boldsymbol\lambda^{k-1}$.\\
Finally, by combining~\eqref{eq:x-subproblem-monotonicity}-\eqref{eq:dual-cross-term-identity}, we obtain
\begin{equation}
\begin{aligned}
0
&\le
\frac{\rho}{2\eta_1}
\|\mathbf x^{k+1}-\mathbf x^k\|_{\mathbf A^\top\mathbf A}^2
-\mu_{F^0}\|\mathbf x^{k+1}-\mathbf x^k\|^2\\
&\quad
+\left(\omega_{H^0}+\rho\eta_1\right)
\|\mathbf B\mathbf z^{k+1}-\mathbf B\mathbf z^k\|^2 \\
&\quad
+\rho\eta_1\|\mathbf B\mathbf z^k-\mathbf B\mathbf z^{k-1}\|^2\\
&\quad
-\frac{1+\tau}{2\rho}
\|\boldsymbol\lambda^{k+1}-\boldsymbol\lambda^k\|^2
+\frac{1-\tau}{2\rho}
\|\boldsymbol\lambda^k-\boldsymbol\lambda^{k-1}\|^2 \\
&\quad
-\frac{\gamma}{2}
\|\mathbf z^{k+1}-\mathbf z^k\|_{\mathbf Q}^2
+\frac{\gamma}{2}
\|\mathbf z^k-\mathbf z^{k-1}\|_{\mathbf Q}^2\\
&\quad
-\frac{1-\tau}{2\rho}
\left\|(\boldsymbol\lambda^{k+1}-\boldsymbol\lambda^k)-(\boldsymbol\lambda^k-\boldsymbol\lambda^{k-1})\right\|^2\\
&\quad
-\frac{\gamma}{2}
\left\|(\mathbf z^{k+1}-\mathbf z^k)-(\mathbf z^k-\mathbf z^{k-1})\right\|_{\mathbf Q}^2 .
\end{aligned}
\label{eq:combined-upper-bound}
\end{equation}
Since \(0<\tau<1\) and \(\gamma\ge0\), then the last two  terms in~\eqref{eq:combined-upper-bound} are non-positive. Hence, after rearrangement, we have
\[
\begin{aligned}
&\frac{\rho}{2\eta_1}
\|\mathbf x^{k+1}-\mathbf x^k\|_{\mathbf A^\top\mathbf A}^2
-\mu_{F^0}\|\mathbf x^{k+1}-\mathbf x^k\|^2\\
&
+\|\mathbf z^{k+1}-\mathbf z^k\|_{\mathbf E_\mathbf z}^2
-\frac{\tau}{\rho}
\|\boldsymbol\lambda^{k+1}-\boldsymbol\lambda^k\|^2 \\
\ge&
\|\mathbf z^{k+1}-\mathbf z^k\|_{\mathbf D_\mathbf z}^2
-\|\mathbf z^k-\mathbf z^{k-1}\|_{\mathbf D_\mathbf z}^2 \\
&
+\frac{1-\tau}{2\rho}
\|\boldsymbol\lambda^{k+1}-\boldsymbol\lambda^k\|^2
-\frac{1-\tau}{2\rho}
\|\boldsymbol\lambda^k-\boldsymbol\lambda^{k-1}\|^2 .
\end{aligned}
\]
where
\[
\mathbf E_{\mathbf z}:=
\left(
\omega_{H^0}
+2\rho\eta_1
\right)\mathbf B^\top\mathbf B,
\quad
\mathbf D_{\mathbf z}:=
\rho\eta_1
\mathbf B^\top\mathbf B
+\frac{\gamma}{2}\mathbf Q .
\]
we therefore complete the proof.

\section{Proof of Proposition~\ref{prop:one_step_descent_AL}}
\label{app:proof_one_step_descent_AL}
We prove the one-step descent estimate by decomposing the change of the augmented Lagrangian into the contributions of the \(\mathbf x\)-update, the \(\mathbf z\)-update, and the dual update.
\paragraph{\(\mathbf x\)-update}
We first estimate the change of the augmented Lagrangian with respect to the \(\mathbf x\)-block while keeping \((\mathbf z^k,\boldsymbol\lambda^k)\) fixed. We have
\begin{equation}
\begin{aligned}
&\mathcal L_{\rho,\tau}(\mathbf x^{k+1},\mathbf z^k,\boldsymbol\lambda^k)-\mathcal L_{\rho,\tau}(\mathbf x^k,\mathbf z^k,\boldsymbol\lambda^k) \\
&=F(\mathbf x^{k+1})-F(\mathbf x^k)
+\left\langle (1-\tau)\boldsymbol\lambda^k,\mathbf A\mathbf x^{k+1}-\mathbf A\mathbf x^k\right\rangle\\
&\quad
+\frac{\rho}{2}\|\mathbf A\mathbf x^{k+1}-\mathbf B\mathbf z^k\|^2
-\frac{\rho}{2}\|\mathbf A\mathbf x^k-\mathbf B\mathbf z^k\|^2 \\
&=F^0(\mathbf x^{k+1})-F^0(\mathbf x^k)
+\left\langle (1-\tau)\boldsymbol\lambda^k,\mathbf A\mathbf x^{k+1}-\mathbf A\mathbf x^k\right\rangle\\
&\quad
+\frac{\rho}{2}\|\mathbf A\mathbf x^{k+1}-\mathbf B\mathbf z^k\|^2
-\frac{\rho}{2}\|\mathbf A\mathbf x^k-\mathbf B\mathbf z^k\|^2 \\
&\overset{\eqref{eq:first_order_condition_x}}{=}F^0(\mathbf x^{k+1})-F^0(\mathbf x^k)
-\left\langle \nabla F^0(\mathbf x^{k+1}),\mathbf x^{k+1}-\mathbf x^k\right\rangle
\\
&
-\rho\left\langle
\mathbf A\mathbf x^{k+1}-\mathbf B\mathbf z^k,\mathbf A\mathbf x^{k+1}-\mathbf A\mathbf x^k
\right\rangle +\frac{\rho}{2}\|\mathbf A\mathbf x^{k+1}-\mathbf B\mathbf z^k\|^2\\
&\quad
-\frac{\rho}{2}\|\mathbf A\mathbf x^k-\mathbf B\mathbf z^k\|^2
-\left\langle \boldsymbol\nu^{F^1}_{\mathbf x^{k+1}},\mathbf x^{k+1}-\mathbf x^k\right\rangle \\
&\le
-\frac{\mu_{F^0}}{2}\|\mathbf x^{k+1}-\mathbf x^k\|^2
-\frac{\rho}{2}\|\mathbf A\mathbf x^{k+1}-\mathbf A\mathbf x^k\|^2 \\
&=
-\frac{1}{2}
\|\mathbf x^{k+1}-\mathbf x^k\|^2_{\mu_{F^0}\mathbf I+\rho\mathbf A^\top\mathbf A}.
\end{aligned}
\label{eq:ALM-x-update-descent}
\end{equation}

The second equality follows from \(F=F^0+F^1\) and \(F^1=\iota_{\Delta_U}\). Indeed, since \(\boldsymbol\nu^{F^1}_{\mathbf x^{k+1}}\in\partial F^1(\mathbf x^{k+1})\), we have \(\partial F^1(\mathbf x^{k+1})\neq\emptyset\), and hence \(\mathbf x^{k+1}\in\Delta_U\). Similarly, \(\mathbf x^k\in\Delta_U\). Thus,
\(F^1(\mathbf x^{k+1})=F^1(\mathbf x^k)=0\).
The inequality in~\eqref{eq:ALM-x-update-descent} then follows directly from Remark~\ref{rem:strong-convexity} applied to \(F^0\), the identity
\(
\frac{1}{2}\|\mathbf a\|^2-\frac{1}{2}\|\mathbf a-\mathbf b\|^2
=\langle \mathbf a,\mathbf b\rangle-\frac{1}{2}\|\mathbf b\|^2,
\)
with \(\mathbf a=\mathbf A\mathbf x^{k+1}-\mathbf B\mathbf z^k\) and \(\mathbf b=\mathbf A\mathbf x^{k+1}-\mathbf A\mathbf x^k\), and the convex subdifferential inequality of \(F^1\):
\[
\begin{aligned}
&-\left\langle \boldsymbol\nu^{F^1}_{\mathbf x^{k+1}},\mathbf x^{k+1}-\mathbf x^k\right\rangle
=
\left\langle \boldsymbol\nu^{F^1}_{\mathbf x^{k+1}},\mathbf x^k-\mathbf x^{k+1}\right\rangle\\
&\le F^1(\mathbf x^k)-F^1(\mathbf x^{k+1})=0.
\end{aligned}
\]

\paragraph{\(\mathbf z\)-update}
We next estimate the change with respect to the \(\mathbf z\)-block while fixing \((\mathbf x^{k+1},\boldsymbol\lambda^k)\). Using the first-order condition of the \(\mathbf z\)-subproblem, we obtain
\begin{equation}
\begin{aligned}
&\mathcal L_{\rho,\tau}(\mathbf x^{k+1},\mathbf z^{k+1},\boldsymbol\lambda^k)
-\mathcal L_{\rho,\tau}(\mathbf x^{k+1},\mathbf z^k,\boldsymbol\lambda^k) \\
&=H^0(\mathbf z^{k+1})-H^0(\mathbf z^k)+H^1(\mathbf z^{k+1})-H^1(\mathbf z^k)\\
&\quad
-\left\langle (1-\tau)\mathbf B^\top\boldsymbol\lambda^k,\mathbf z^{k+1}-\mathbf z^k\right\rangle+\frac{\rho}{2}\|\mathbf A\mathbf x^{k+1}-\mathbf B\mathbf z^{k+1}\|^2 \\
&\quad
-\frac{\rho}{2}\|\mathbf A\mathbf x^{k+1}-\mathbf B\mathbf z^k\|^2 \\
&\overset{\eqref{eq:first_order_condition_z}}{=}H^0(\mathbf z^{k+1})-H^0(\mathbf z^k)-\left\langle \nabla H^0(\mathbf z^{k+1}),\mathbf z^{k+1}-\mathbf z^k\right\rangle \\
&\quad
+H^1(\mathbf z^{k+1})-H^1(\mathbf z^k)
-\left\langle \boldsymbol\nu^{H^1}_{\mathbf z^{k+1}},\mathbf z^{k+1}-\mathbf z^k\right\rangle \\
&\quad
+\rho\left\langle
\mathbf A\mathbf x^{k+1}-\mathbf B\mathbf z^{k+1},\mathbf B\mathbf z^{k+1}-\mathbf B\mathbf z^k
\right\rangle
-\gamma\|\mathbf z^{k+1}-\mathbf z^k\|_{\mathbf Q}^2 \\
&\quad
+\frac{\rho}{2}\|\mathbf A\mathbf x^{k+1}-\mathbf B\mathbf z^{k+1}\|^2
-\frac{\rho}{2}\|\mathbf A\mathbf x^{k+1}-\mathbf B\mathbf z^k\|^2 \\
&\le \frac{\omega_{H^0}}{2}\|\mathbf B\mathbf z^{k+1}-\mathbf B\mathbf z^k\|^2
-\frac{\rho}{2}\|\mathbf B\mathbf z^{k+1}-\mathbf B\mathbf z^k\|^2\\
&\quad
-\gamma\|\mathbf z^{k+1}-\mathbf z^k\|_{\mathbf Q}^2 \\
&= -\frac{1}{2}
\|\mathbf z^{k+1}-\mathbf z^k\|^2_{
(\rho-\omega_{H^0})\mathbf B^\top \mathbf B+2\gamma \mathbf Q}.
\end{aligned}
\label{eq:ALM-z-update-descent}
\end{equation}
The inequality in~\eqref{eq:ALM-z-update-descent} is obtained by using the \(\omega_{H^0}\)-restricted weak convexity of \(H^0\) with respect to \(\mathbf B\), the convexity of \(H^1\), and the identity
\(
\frac{1}{2}\|\mathbf a\|^2-\frac{1}{2}\|\mathbf a+\mathbf b\|^2
=-\langle \mathbf a,\mathbf b\rangle-\frac{1}{2}\|\mathbf b\|^2,
\)
with \(\mathbf a=\mathbf A\mathbf x^{k+1}-\mathbf B\mathbf z^{k+1}\) and \(\mathbf b=\mathbf B\mathbf z^{k+1}-\mathbf B\mathbf z^k\).

\paragraph{\(\boldsymbol\lambda\)-update}
It remains to quantify the change induced by the dual update. By~\eqref{eq:dual-update}, we have
\begin{equation}
\begin{aligned}
&\mathcal L_{\rho,\tau}(\mathbf x^{k+1},\mathbf z^{k+1},\boldsymbol\lambda^{k+1})
-\mathcal L_{\rho,\tau}(\mathbf x^{k+1},\mathbf z^{k+1},\boldsymbol\lambda^k) \\
&\overset{\eqref{eq:dual-update}}{=}\frac{1-\tau}{\rho}
\left\langle \boldsymbol\lambda^{k+1}-\boldsymbol\lambda^k,
(\boldsymbol\lambda^{k+1}-\boldsymbol\lambda^k)+\tau\boldsymbol\lambda^k
\right\rangle \\
&=\frac{(1-\tau)(2-\tau)}{2\rho}\|\boldsymbol\lambda^{k+1}-\boldsymbol\lambda^k\|^2\\
&\quad
+\frac{\tau(1-\tau)}{2\rho}
\left(\|\boldsymbol\lambda^{k+1}\|^2-\|\boldsymbol\lambda^k\|^2\right).
\end{aligned}
\label{eq:ALM-lambda-update}
\end{equation}
In the last equality, we used
\[
2\left\langle \boldsymbol\lambda^{k+1}-\boldsymbol\lambda^k,\boldsymbol\lambda^k\right\rangle
=
\|\boldsymbol\lambda^{k+1}\|^2-\|\boldsymbol\lambda^k\|^2
-\|\boldsymbol\lambda^{k+1}-\boldsymbol\lambda^k\|^2.
\]
Summing~\eqref{eq:ALM-x-update-descent}, \eqref{eq:ALM-z-update-descent}, and~\eqref{eq:ALM-lambda-update}, then we complete the proof.

\section{Proof of~Lemma~\ref{lem:sufficient_descent_lyapunov}}
\label{app:proof_sufficient_descent_lyapunov}

Based on Proposition~\ref{prop:combined_descent_rearranged} and
Proposition~\ref{prop:one_step_descent_AL}, we have the following relationship:
\begin{equation}
\begin{aligned}
&\mathcal P(\mathbf w^{k+1})-\mathcal P(\mathbf w^k) \\
\le&
-\frac{1}{2}
\left\|\mathbf x^{k+1}-\mathbf x^k\right\|^2_{\mathbf M_\mathbf x}
-\frac{1}{2}
\left\|\mathbf z^{k+1}-\mathbf z^k\right\|^2_{\mathbf M_\mathbf z}\\
&-a_\lambda
\|\boldsymbol\lambda^{k+1}-\boldsymbol\lambda^k\|^2 ,
\end{aligned}
\label{eq:Lyapunov-descent-combined}
\end{equation}
where
\[
\begin{aligned}
&\mathbf E_\mathbf z:=
\left(
\omega_{H^0}
+2\rho\eta_1
\right)\mathbf B^\top\mathbf B,\\
&\mathbf M_\mathbf x
:=
\left(1+2d\right)\mu_{F^0}\mathbf I+\left(1-\frac{d}{\eta_1}\right)\rho\mathbf A^\top\mathbf A,\\
&\mathbf M_\mathbf z:=
(\rho-\omega_{H^0})\mathbf B^\top\mathbf B
+2\gamma\mathbf Q
-2d\mathbf E_\mathbf z,\\
&a_{\boldsymbol\lambda}:=
\frac{d\tau}{\rho}
-\frac{(1-\tau)(2-\tau)}{2\rho}.
\end{aligned}
\]
By the parameter conditions in Lemma~\ref{lem:sufficient_descent_lyapunov}, we can easily verify that
\(
\mathbf M_\mathbf x\succ0,\, \mathbf M_\mathbf z\succ0,\, a_{\boldsymbol\lambda}>0.
\)
Therefore, from \eqref{eq:Lyapunov-descent-combined},
\[
\begin{aligned}
&\mathcal P(\mathbf w^{k+1})-\mathcal P(\mathbf w^k)\\
\le&
-C_1\left(
\|\mathbf x^{k+1}-\mathbf x^k\|^2
+\|\mathbf z^{k+1}-\mathbf z^k\|^2
+\|\boldsymbol\lambda^{k+1}-\boldsymbol\lambda^k\|^2
\right),
\end{aligned}
\]
where
\[
C_1:=
\min\left\{
\frac{1}{2}\lambda_{\min}(\mathbf M_\mathbf x),
\frac{1}{2}\lambda_{\min}(\mathbf M_\mathbf z),
a_{\boldsymbol\lambda}
\right\}>0.
\]
This proves the desired sufficient descent estimate.

\section{Proof of Lemma~\ref{lem:lower_boundedness_lyapunov}}
\label{app:proof_lower_boundedness_lyapunov}

Since \(F\) and \(H\) are lower bounded on the corresponding probability sets, there exist constants \(\underline F\) and \(\underline H\) such that
\[
F(\mathbf x^k)\ge \underline F,\qquad
H(\mathbf z^k)\ge \underline H,\qquad \forall k\in \mathbb N .
\]
Let
\(
\underline{\mathcal P}:=\underline F+\underline H .
\)
Under the parameter conditions of Lemma~\ref{lem:sufficient_descent_lyapunov}, the additional quadratic terms in \(\mathcal P\) are nonnegative. Hence, by the definition of \(\mathcal P\), for every \(k\in\mathbb N\),
\[
\begin{aligned}
\mathcal P(\mathbf w^{k+1})
&\ge
\underline{\mathcal P}
+(1-\tau)\left\langle \boldsymbol\lambda^{k+1},
\mathbf A\mathbf x^{k+1}-\mathbf B\mathbf z^{k+1}\right\rangle\\
&\quad
-\frac{\tau(1-\tau)}{2\rho}\|\boldsymbol\lambda^{k+1}\|^2 .
\end{aligned}
\]
Using the dual update~\eqref{eq:dual-update}, we have
\[
\begin{aligned}
&(1-\tau)\left\langle \boldsymbol\lambda^{k+1},
\mathbf A\mathbf x^{k+1}-\mathbf B\mathbf z^{k+1}\right\rangle
-\frac{\tau(1-\tau)}{2\rho}\|\boldsymbol\lambda^{k+1}\|^2  \\
=&
\frac{(1-\tau)(2-\tau)}{2\rho}\|\boldsymbol\lambda^{k+1}\|^2
-\frac{(1-\tau)^2}{\rho}
\left\langle \boldsymbol\lambda^{k+1},\boldsymbol\lambda^k\right\rangle  \\
=&
\frac{1-\tau}{2\rho}\|\boldsymbol\lambda^{k+1}\|^2
-\frac{(1-\tau)^2}{2\rho}\|\boldsymbol\lambda^k\|^2\\
&
+\frac{(1-\tau)^2}{2\rho}
\|\boldsymbol\lambda^{k+1}-\boldsymbol\lambda^k\|^2 .
\end{aligned}
\]
Consequently, by dropping the last nonnegative term, we obtain
\begin{equation}
\mathcal P(\mathbf w^{k+1})
\ge
\underline{\mathcal P}
+\frac{1-\tau}{2\rho}\|\boldsymbol\lambda^{k+1}\|^2
-\frac{(1-\tau)^2}{2\rho}\|\boldsymbol\lambda^k\|^2 .
\label{eq:app-P-lower-preliminary}
\end{equation}
Choose 
\[
M:=
\max\left\{
\|\boldsymbol\lambda^0\|^2,\,
\frac{2\rho}{\tau(1-\tau)}
\left(\mathcal P(\mathbf w^0)-\underline{\mathcal P}\right)
\right\}<+\infty.
\]
We next prove by induction that
\[
\|\boldsymbol\lambda^k\|^2\le M,\qquad \forall \, k\in\mathbb N .
\]
The claim is trivial for \(k=0\). Let \(K\in\mathbb N\) be fixed and assume that
\[
\|\boldsymbol\lambda^k\|^2\le M,\qquad k=0,1,\ldots,K .
\]
We show that the same bound holds for \(k=K+1\).\\
If
\(
\|\boldsymbol\lambda^{K+1}\|^2\le \|\boldsymbol\lambda^K\|^2,
\)
then
\(
\|\boldsymbol\lambda^{K+1}\|^2\le \|\boldsymbol\lambda^K\|^2\le M .
\)\\
If instead
\(
\|\boldsymbol\lambda^{K+1}\|^2> \|\boldsymbol\lambda^K\|^2,
\)
then, since \(0<\tau<1\), we have \(0<1-\tau<1\). Hence, by~\eqref{eq:app-P-lower-preliminary} with \(k=K\),
\[
\begin{aligned}
\mathcal P(\mathbf w^{K+1})
&\ge
\underline{\mathcal P}
+\frac{1-\tau}{2\rho}\|\boldsymbol\lambda^{K+1}\|^2
-\frac{(1-\tau)^2}{2\rho}\|\boldsymbol\lambda^K\|^2\\
&\ge
\underline{\mathcal P}
+\frac{1-\tau}{2\rho}\|\boldsymbol\lambda^{K+1}\|^2
-\frac{(1-\tau)^2}{2\rho}\|\boldsymbol\lambda^{K+1}\|^2\\
&=
\underline{\mathcal P}
+\frac{\tau(1-\tau)}{2\rho}\|\boldsymbol\lambda^{K+1}\|^2 .
\end{aligned}
\]
By the nonincreasing property of \(\{\mathcal P(\mathbf w^k)\}_{k\in\mathbb N}\) established in Lemma~\ref{lem:sufficient_descent_lyapunov}, we have
\[
\begin{aligned}
\|\boldsymbol\lambda^{K+1}\|^2
&\le
\frac{2\rho}{\tau(1-\tau)}
\left(\mathcal P(\mathbf w^{K+1})-\underline{\mathcal P}\right)\\
&\le
\frac{2\rho}{\tau(1-\tau)}
\left(\mathcal P(\mathbf w^0)-\underline{\mathcal P}\right)
\le M .
\end{aligned}
\]
Thus, \(\|\boldsymbol\lambda^{K+1}\|^2\le M\) in both cases, and the induction is complete. Using this bound in~\eqref{eq:app-P-lower-preliminary}, we obtain, for all \(k\in\mathbb N\),
\[
\mathcal P(\mathbf w^{k+1})
\ge
\underline{\mathcal P}
-\frac{(1-\tau)^2}{2\rho}M
>-\infty .
\]
Therefore, the Lyapunov sequence \(\{\mathcal P(\mathbf w^k)\}_{k\in \mathbb N}\) is bounded from below. This completes the proof.

\section{Proof of Theorem~\ref{thm:subsequential_convergence}}
\label{app:proof_subsequential_convergence}

We first sum the sufficient descent inequality and use the lower boundedness of the Lyapunov sequence to show that the successive changes of the primal and dual variables are square summable and hence converge to zero. Summing
\eqref{eq:sufficient-descent-C1} from \(k=0\) to \(K-1\), we obtain
\begin{equation}
\begin{aligned}
C_1\sum_{k=0}^{K-1}
\Big(
&\|\mathbf x^{k+1}-\mathbf x^k\|^2
+\|\mathbf z^{k+1}-\mathbf z^k\|^2 \nonumber \\
&+\|\boldsymbol\lambda^{k+1}-\boldsymbol\lambda^k\|^2
\Big)
\le
\mathcal P(\mathbf w^0)-\mathcal P(\mathbf w^K).
\end{aligned}
\label{eq:app-thm-telescoping}
\end{equation}
Letting \(K\to\infty\) and using the lower boundedness of \(\{\mathcal P(\mathbf w^k)\}_{k\in \mathbb N}\), we obtain
\[  
\begin{aligned}
\sum_{k=0}^{\infty}
\Big(
\|\mathbf x^{k+1}-\mathbf x^k&\|^2
+\|\mathbf z^{k+1}-\mathbf z^k\|^2\\
&+\|\boldsymbol\lambda^{k+1}-\boldsymbol\lambda^k\|^2
\Big)
<+\infty.
\end{aligned}
\]
Consequently,
\begin{equation}
\mathbf x^{k+1}-\mathbf x^k\to0,\qquad
\mathbf z^{k+1}-\mathbf z^k\to0,\qquad
\boldsymbol\lambda^{k+1}-\boldsymbol\lambda^k\to0.
\label{eq:app-thm-vanishing-difference}
\end{equation}

Since \(\mathbf x^k\) and \(\mathbf z^k\) remain in the corresponding probability feasible sets, the sequences \(\{\mathbf x^k\}_{k\in\mathbb N}\) and \(\{\mathbf z^k\}_{k\in\mathbb N}\) are bounded. 
In addition, Lemma~\ref{lem:lower_boundedness_lyapunov} shows that \(\{\|\boldsymbol\lambda^k\|\}_{k\in\mathbb N}\) is bounded from above, and hence the sequence \(\{\boldsymbol\lambda^k\}_{k\in\mathbb N}\) is bounded. Thus, \(\{\mathbf u^k\}_{k\in\mathbb N}\) is bounded and has at least one accumulation point. Then there exists a subsequence \(\{\mathbf u^{k_j}\}_{j\in\mathbb N}\) such that
\(
\mathbf u^{k_j}\to \mathbf u^\ast .
\)
By \eqref{eq:app-thm-vanishing-difference}, we also have
\(
\mathbf u^{k_j+1}\to \mathbf u^\ast .
\)

We next verify the stationarity conditions. 
From~\eqref{eq:first_order_conditions}, the first-order conditions of \(\mathbf x\)-subproblem and \(\mathbf z\)-subproblem can be rewritten as
\[
-\rho\mathbf A^\top\mathbf B(\mathbf z^{k+1}-\mathbf z^k)
\in
\nabla F^0(\mathbf x^{k+1})
+\partial F^1(\mathbf x^{k+1})
+\mathbf A^\top\boldsymbol\lambda^{k+1},
\]
\[
-\gamma\mathbf Q(\mathbf z^{k+1}-\mathbf z^k)
\in
\nabla H^0(\mathbf z^{k+1})
+\partial H^1(\mathbf z^{k+1})
-\mathbf B^\top\boldsymbol\lambda^{k+1}.
\]
Setting \(k=k_j\) and passing to the limit as \(j\to\infty\), together
with the vanishing successive differences, the continuity of
\(\nabla F^0\) and \(\nabla H^0\), and the closedness of the limiting
subdifferential, yields
\begin{equation}
0\in
\nabla F^0(\mathbf x^\ast)
+\partial F^1(\mathbf x^\ast)
+\mathbf A^\top\boldsymbol\lambda^\ast,
\label{eq:app-thm-x-stationarity-limit}
\end{equation}
\begin{equation}
0\in
\nabla H^0(\mathbf z^\ast)
+\partial H^1(\mathbf z^\ast)
-\mathbf B^\top\boldsymbol\lambda^\ast .
\label{eq:app-thm-z-stationarity-limit}
\end{equation}

It remains to estimate the feasibility residual.
The dual update~\eqref{eq:dual-update} gives
\[
\mathbf A\mathbf x^{k+1}-\mathbf B\mathbf z^{k+1}
=
\frac{1}{\rho}
(\boldsymbol\lambda^{k+1}-\boldsymbol\lambda^k)
+\frac{\tau}{\rho}\boldsymbol\lambda^k .
\]
Therefore,
\[
\|\mathbf A\mathbf x^{k+1}-\mathbf B\mathbf z^{k+1}\|
\le
\frac{1}{\rho}\|\boldsymbol\lambda^{k+1}-\boldsymbol\lambda^k\|
+\frac{\tau}{\rho}\|\boldsymbol\lambda^k\|.
\]
Taking the limit superior and using
\(\boldsymbol\lambda^{k+1}-\boldsymbol\lambda^k\to0\), we obtain
\[
\limsup_{k\to\infty}\|\mathbf A\mathbf x^{k+1}-\mathbf B\mathbf z^{k+1}\|
\le
\frac{\tau}{\rho}
\limsup_{k\to\infty}\|\boldsymbol\lambda^k\|.
\]
Since \(\mathbf u^{k_j+1}\to\mathbf u^\ast\), it follows that
\[
\mathbf A\mathbf x^{k_j+1}-\mathbf B\mathbf z^{k_j+1}
\to
\mathbf A\mathbf x^\ast-\mathbf B\mathbf z^\ast .
\]
Therefore,
\begin{equation}
\|\mathbf A\mathbf x^\ast-\mathbf B\mathbf z^\ast\|
\le
\frac{\tau}{\rho}
\limsup_{k\to\infty}\|\boldsymbol\lambda^k\|.
\label{eq:app-thm-feasibility-residual-bound}
\end{equation}
The stationarity conditions
\eqref{eq:app-thm-x-stationarity-limit} and
\eqref{eq:app-thm-z-stationarity-limit}, together with the feasibility
bound \eqref{eq:app-thm-feasibility-residual-bound}, constitute the
\(\epsilon\)-KKT conditions at \(\mathbf u^\ast\), where
\[
\epsilon
:=
\frac{\tau}{\rho}
\limsup_{k\to\infty}\|\boldsymbol\lambda^k\|.
\]
Hence, \(\mathbf u^\ast\) is an \(\epsilon\)-KKT point. This completes
the proof.
\section{Proof of Lemma~\ref{lem:subgradient_bound}}
\label{app:proof_subgradient_bound}
Let
\(
\mathbf w
:=
\bigl(
\mathbf x,
\mathbf z,
\boldsymbol\lambda,
\mathbf z',
\boldsymbol\lambda'
\bigr)
\)
denote the argument of \(\mathcal P\). We then estimate the components of
\(\partial_{\mathbf w}\mathcal P(\mathbf w^{k+1})\) separately.
\paragraph{\(\mathbf x\)-component}

Using the first-order condition~\eqref{eq:first_order_condition_x}, we obtain
\begin{equation}
\begin{aligned}
&\operatorname{dist}\!\left(\mathbf 0,\partial_{\mathbf x}\mathcal P(\mathbf w^{k+1})\right)\\
&\le
\left\|
\nabla F^0(\mathbf x^{k+1})
+\boldsymbol\nu_{\mathbf x^{k+1}}^{F^1}
+(1-\tau)\mathbf A^\top\boldsymbol\lambda^{k+1}
\right.\\
&\left.\quad
+\rho\mathbf A^\top(\mathbf A\mathbf x^{k+1}-\mathbf B\mathbf z^{k+1})
\right\| \\
&=
\left\|
(1-\tau)\mathbf A^\top(\boldsymbol\lambda^{k+1}-\boldsymbol\lambda^k)
+\rho\mathbf A^\top\mathbf B(\mathbf z^k-\mathbf z^{k+1})
\right\| \\
&\le
(1-\tau)\|\mathbf A^\top\|\,\|\boldsymbol\lambda^{k+1}-\boldsymbol\lambda^k\|
+\rho\|\mathbf A^\top\mathbf B\|\,\|\mathbf z^{k+1}-\mathbf z^{k}\|.
\end{aligned}
\label{eq:app-subgradient-bound-x}
\end{equation}

\paragraph{\(\mathbf z\)-component}
Similarly, by the first-order condition~\eqref{eq:first_order_condition_z}, we have
\begin{equation}
\begin{aligned}
&\operatorname{dist}\!\left(\mathbf 0,\partial_{\mathbf z}\mathcal P(\mathbf w^{k+1})\right)\\
\le&
\Big\|
\nabla H^0(\mathbf z^{k+1})
+\boldsymbol\nu_{\mathbf z^{k+1}}^{H^1}
-(1-\tau)\mathbf B^\top\boldsymbol\lambda^{k+1}\\
&
-\rho\mathbf B^\top(\mathbf A\mathbf x^{k+1}-\mathbf B\mathbf z^{k+1})
+2d\rho\eta_1\mathbf B^\top\mathbf B(\boldsymbol\lambda^{k+1}-\boldsymbol\lambda^k)\\
&
+d\gamma\mathbf Q(\boldsymbol\lambda^{k+1}-\boldsymbol\lambda^k)
\Big\| \\
=&
\Big\|
(1-\tau)\mathbf B^\top(\boldsymbol\lambda^k-\boldsymbol\lambda^{k+1})
+2d\rho\eta_1\mathbf B^\top\mathbf B(\mathbf z^{k+1}-\mathbf z^{k})\\
&
+(d-1)\gamma\mathbf Q(\mathbf z^{k+1}-\mathbf z^{k})
\Big\| \\
\le&
(1-\tau)\|\mathbf B^\top\|\,\|\boldsymbol\lambda^{k+1}-\boldsymbol\lambda^k\|\\
&
+\left(
2d\rho\eta_1\|\mathbf B^\top\mathbf B\|
+|d-1|\gamma\|\mathbf Q\|
\right)\|\mathbf z^{k+1}-\mathbf z^{k}\|.
\end{aligned}
\label{eq:app-subgradient-bound-z}
\end{equation}

\paragraph{\(\boldsymbol\lambda\)-component}
From the dual update~\eqref{eq:dual-update}, we have
\begin{equation}
\begin{aligned}
&\operatorname{dist}\!\left(\mathbf 0,\partial_{\boldsymbol\lambda}\mathcal P(\mathbf w^{k+1})\right)\\
=&
\Bigg\|
(1-\tau)(\mathbf A\mathbf x^{k+1}-\mathbf B\mathbf z^{k+1})+d\frac{1-\tau}{\rho}(\boldsymbol\lambda^{k+1}-\boldsymbol\lambda^k)\\
&\qquad
-\frac{\tau(1-\tau)}{\rho}\boldsymbol\lambda^{k+1}
\Bigg\| \\
=&
\frac{(1-\tau)(d+1-\tau)}{\rho}
\|\boldsymbol\lambda^{k+1}-\boldsymbol\lambda^k\|.
\end{aligned}
\label{eq:app-subgradient-bound-lambda}
\end{equation}

\paragraph{Auxiliary components}
For the auxiliary variable \(\mathbf z'\), we have
\begin{equation}
\begin{aligned}
&\operatorname{dist}\!\left(\mathbf 0,\partial_{\mathbf z'}\mathcal P(\mathbf w^{k+1})\right)\\
&=
\left\|
-2d\rho\eta_1\mathbf B^\top\mathbf B(\mathbf z^{k+1}-\mathbf z^{k})
-d\gamma\mathbf Q(\mathbf z^{k+1}-\mathbf z^{k})
\right\| \\
&\le
\left(
2d\rho\eta_1\|\mathbf B^\top\mathbf B\|
+d\gamma\|\mathbf Q\|
\right)\|\mathbf z^{k+1}-\mathbf z^{k}\|.
\end{aligned}
\label{eq:app-subgradient-bound-zprime}
\end{equation}
For the auxiliary variable \(\boldsymbol\lambda'\), it follows that
\begin{equation}
\operatorname{dist}\!\left(\mathbf 0,\partial_{\boldsymbol\lambda'}\mathcal P(\mathbf w^{k+1})\right)
=
d\frac{1-\tau}{\rho}\|\boldsymbol\lambda^{k+1}-\boldsymbol\lambda^k\|.
\label{eq:app-subgradient-bound-lambdaprime}
\end{equation}

Combining~\eqref{eq:app-subgradient-bound-x}--\eqref{eq:app-subgradient-bound-lambdaprime}, we obtain
\[
\operatorname{dist}\!\left(\mathbf 0,\partial_\mathbf w\mathcal P(\mathbf w^{k+1})\right)
\le
C_{\boldsymbol\lambda}^{\rm sg}\|\boldsymbol\lambda^{k+1}-\boldsymbol\lambda^k\|
+
C_{\mathbf z}^{\rm sg}\|\mathbf z^{k+1}-\mathbf z^{k}\|,
\]
where
\[
\begin{aligned}
&C_{\boldsymbol\lambda}^{\rm sg}:=
(1-\tau)(\|\mathbf A^\top\|+\|\mathbf B^\top\|)
+\frac{(1-\tau)(2d+1-\tau)}{\rho}>0,\\
&C_{\mathbf z}^{\rm sg}:=
\rho\|\mathbf A^\top\mathbf B\|
+4d\rho\eta_1\|\mathbf B^\top\mathbf B\|+(|d-1|+d)\gamma\|\mathbf Q\|>0.
\end{aligned}
\]
Let
\(
C_2:=\sqrt{(C_{\boldsymbol\lambda}^{\rm sg})^2+(C_{\mathbf z}^{\rm sg})^2}.
\)
Then,
\[
\begin{aligned}
&\operatorname{dist}\!\left(\mathbf 0,\partial_\mathbf w\mathcal P(\mathbf w^{k+1})\right)\\
\le
&C_2
\left(
\|\boldsymbol\lambda^{k+1}-\boldsymbol\lambda^k\|^2
+\|\mathbf z^{k+1}-\mathbf z^{k}\|^2
\right)^{1/2}\\
\le
&C_2\left(
\|\mathbf x^{k+1}-\mathbf x^{k}\|^2
+\|\mathbf z^{k+1}-\mathbf z^{k}\|^2
+\|\boldsymbol\lambda^{k+1}-\boldsymbol\lambda^k\|^2
\right)^{1/2}.
\end{aligned}
\]
This completes the proof.

\section{Proof of Lemma~\ref{lem:KL_property}}
\label{app:proof_KL_property}
The proof of this lemma is based mainly on proving the following result: 
\begin{lemma}[Definability of the Lyapunov function \(\mathcal P\)]
\label{lem:lyapunov_definable}
The Lyapunov function \(\mathcal P\) is definable in 
\(
\mathbb R_{\exp}:=(\mathbb R,+,\cdot,<,\exp).
\)
\end{lemma}
The proof of this lemma is provided in Appendix~\ref{app:lyapunov_definability}. 
By this lemma, the Lyapunov function \(\mathcal P\) is \(\mathbb R_{\exp}\)-definable, where \(\mathbb R_{\exp}\) is an o-minimal structure by Wilkie's
theorem~\cite{wilkie1996model}. Moreover, Appendix~\ref{app:proper_lsc_P} shows that \(\mathcal P\) is proper and
lower semicontinuous. Hence, \cite[Theorem~14]{bolte2007clarke} implies that \(\mathcal P\) satisfies the Kurdyka--{\L}ojasiewicz property at every point in \(\operatorname{dom}\partial\mathcal P\). This completes the proof.

\section{Proof of Lemma~\ref{lem:descent_subgradient_recursion}}
\label{app:proof_descent_subgradient_recursion}
The proof consists in combining the sufficient descent inequality, the subgradient bound, and the K{\L} inequality to derive a recursion for the Lyapunov error~\cite{Attouch2013}. 
By Lemmas~\ref{lem:sufficient_descent_lyapunov}
and~\ref{lem:subgradient_bound}, we have
\begin{equation}
\mathcal P(\mathbf w^k)-\mathcal P(\mathbf w^{k+1})
\ge
\frac{C_1}{C_2^2}
\operatorname{dist}^2\!\left(\mathbf 0,\partial_\mathbf w\mathcal P(\mathbf w^{k+1})
\right).
\label{eq:descent-subgradient-estimate}
\end{equation}
Recall that \(\mathbf w^\ast\) is the limit point of \(\{\mathbf w^k\}_{k\in\mathbb N}\). By the convergence result
established above,
\[
\mathbf w^k\to \mathbf w^\ast,
\qquad
\mathcal P(\mathbf w^k)\to\mathcal P(\mathbf w^\ast),
\]
and \(\mathbf w^\ast\) is a critical point of \(\mathcal P\). Hence, by
Lemma~\ref{lem:KL_property}, there exists \(k_0\in\mathbb N\) such that
\[
\operatorname{dist}\!\left(\mathbf 0,\partial_\mathbf w\mathcal P(\mathbf w^{k+1})
\right)
\ge
C_3 e_{k+1}^{\theta},
\qquad
\forall \, k\in\mathbb N,\ k\ge k_0.
\]
Combining this inequality with
\eqref{eq:descent-subgradient-estimate} yields
\[
\begin{aligned}
e_k-e_{k+1}
=
\mathcal P(\mathbf w^k)-\mathcal P(\mathbf w^{k+1})
\ge
\frac{C_1C_3^2}{C_2^2}e_{k+1}^{2\theta}.
\end{aligned}
\]
Therefore, defining
\(
\bar C:=\frac{C_1C_3^2}{C_2^2}>0,
\)
we obtain
\[
e_k-e_{k+1}
\ge
\bar C e_{k+1}^{2\theta},
\qquad
\forall \, k\in\mathbb N,\ k\ge k_0.
\]
This completes the proof.

\section{Proof of Theorem~\ref{thm:rate_lyapunov_error}}
\label{app:proof_rate_lyapunov_error}
Inspired by~\cite{Attouch2013,Frankel2015,Yashtini2021}, we provide an error decaying recursion for different values of \(\theta\).
\paragraph{\(\theta=0\)}
then Lemma~\ref{lem:descent_subgradient_recursion} gives \(e_k-e_{k+1}\ge\bar C\) whenever
\(e_{k+1}>0\). Since \(\{e_k\}_{k\in\mathbb N}\) is nonnegative, this cannot occur
infinitely often. Hence, \(e_k=0\) for all sufficiently large \(k\).
\paragraph{\(\theta\in(0,1/2]\)}
We first consider \(\theta\in(0,1/2)\). If \(e_{k_0}=0\), then the
nonnegativity and monotonicity of \(\{e_k\}_{k\in\mathbb N}\) imply that \(e_k=0\) for all
\(k\in\mathbb N,\ k\ge k_0\), and the conclusion follows immediately. Hence, assume
\(e_{k_0}>0\). Since \(\{e_k\}_{k\in\mathbb N}\) is nonincreasing and \(2\theta-1<0\), we have
\[
e_{k+1}^{2\theta}
=
e_{k+1}e_{k+1}^{2\theta-1}
\ge
e_{k+1}e_{k_0}^{2\theta-1},
\qquad \forall \, k\in\mathbb N,\ k\ge k_0.
\]
Combining this inequality with \eqref{eq:KL-recursion-lemma} yields
\[
e_{k+1}
\le
\frac{1}{1+\bar C e_{k_0}^{2\theta-1}}\,e_k.
\]
Iterating the above inequality gives
\[
e_k
\le
e_{k_0}
\left(1+\bar C e_{k_0}^{2\theta-1}\right)^{-(k-k_0)},
\qquad \forall \, k\in\mathbb N,\ k\ge k_0.
\]
Next, let \(\theta=1/2\). In this case,
\eqref{eq:KL-recursion-lemma} reduces to
\[
e_{k+1}\le\frac{1}{1+\bar C}\,e_k.
\]
Hence,
\[
e_k
\le
e_{k_0}(1+\bar C)^{-(k-k_0)},
\qquad \forall \, k\in\mathbb N,\ k\ge k_0.
\]
Therefore, for every \(\theta\in(0,1/2]\),
\[
e_k
\le
e_{k_0}
\left(1+\bar C e_{k_0}^{2\theta-1}\right)^{-(k-k_0)},
\qquad \forall \, k\in\mathbb N,\ k\ge k_0.
\]
\paragraph{\(\theta\in(1/2,1)\)}
If \(e_k=0\) for some sufficiently large \(k\), then the nonnegativity and monotonicity of \(\{e_k\}_{k\in\mathbb N}\) imply that \(e_j=0\) for all \(j\ge k\), and the conclusion follows immediately. Hence, we only need to consider the case \(e_k>0\) for all \(k\ge k_0\). Since \(\{e_k\}_{k\in\mathbb N}\) is nonincreasing and \(2\theta-1>0\), we have
\[
0<
\bar C e_{k+1}^{2\theta-1}
\le
\bar C e_{k_0}^{2\theta-1},
\qquad
\forall \, k\in\mathbb N,\ k\ge k_0.
\]
Moreover, \eqref{eq:KL-recursion-lemma} yields
\[
\begin{aligned}
e_{k+1}^{1-2\theta}-e_k^{1-2\theta}
&\ge
e_{k+1}^{1-2\theta}
\left[
1-
\left(
1+\bar C e_{k+1}^{2\theta-1}
\right)^{1-2\theta}
\right].
\end{aligned}
\]
For every
\(
s\in
\left[
0,\bar C e_{k_0}^{2\theta-1}
\right],
\)
the Lagrange mean value theorem applied to
\(\psi(s)=(1+s)^{1-2\theta}\) gives
\[
1-(1+s)^{1-2\theta}
\ge
\frac{2\theta-1}
{\left(1+\bar C e_{k_0}^{2\theta-1}\right)^{2\theta}}
\,s.
\]
Taking \(s=\bar C e_{k+1}^{2\theta-1}\), we obtain
\begin{equation}
e_{k+1}^{1-2\theta}-e_k^{1-2\theta}
\ge
\frac{(2\theta-1)\bar C}
{\left(1+\bar C e_{k_0}^{2\theta-1}\right)^{2\theta}}.
\label{eq:inverse-power-increment}
\end{equation}
Defining
\[
\mu
:=
\frac{(2\theta-1)\bar C}
{\left(1+\bar C e_{k_0}^{2\theta-1}\right)^{2\theta}}
>0,
\]
and summing~\eqref{eq:inverse-power-increment} from \(k_0\) to \(k-1\),
we obtain
\[
e_k^{1-2\theta}
\ge
e_{k_0}^{1-2\theta}
+\mu(k-k_0).
\]
Since \(1-2\theta<0\), it follows that
\[
e_k
\le
\left(
\mu(k-k_0)+e_{k_0}^{1-2\theta}
\right)^{-\frac{1}{2\theta-1}},
\qquad
\forall \, k\in\mathbb N,\ k\ge k_0.
\]
This completes the proof.

\section{Proof of Theorem~\ref{thm:rate_iterates}}
\label{app:proof_rate_iterates}

Let
\(
S_{k+1}
:=
\|\mathbf x^{k+1}-\mathbf x^k\|
+\|\mathbf z^{k+1}-\mathbf z^k\|
+\|\boldsymbol\lambda^{k+1}-\boldsymbol\lambda^k\|.
\)
Since \(\mathbf u^k\to \mathbf u^\ast\), the triangle inequality gives, for every
\(k\ge k_0\),
\begin{equation}
\|\mathbf u^k-\mathbf u^\ast\|
\le
\sum_{p\ge k}\|\mathbf u^{p+1}-\mathbf u^p\|
\le
\sum_{p\ge k}S_{p+1}.
\label{eq:uk-tail-bound}
\end{equation}
Let
\(
\psi(s):=\frac{1}{C_3(1-\theta)}s^{1-\theta}.
\)
Since \(\psi\) is concave, we have
\[
\psi(e_k)-\psi(e_{k+1})
\geq
\psi'(e_k)(e_k-e_{k+1}).
\]
Moreover, \eqref{eq:sufficient-descent-C1} can be written as
\begin{equation}
\begin{aligned}
&e_k-e_{k+1}\\
&\geq 
C_1\left(
\|\mathbf x^{k+1}-\mathbf x^k\|^2
+\|\mathbf z^{k+1}-\mathbf z^k\|^2
+\|\boldsymbol\lambda^{k+1}-\boldsymbol\lambda^k\|^2
\right)\\
&\geq
\frac{C_1}{3}S_{k+1}^2.
\end{aligned}
\label{eq:step-error-bound}
\end{equation}
where the last inequality follows from the Cauchy--Schwarz inequality \(\left(\sum_{i=1}^{n} a_i\right)^2\leq n\sum_{i=1}^{n}a_i^2\), applied with \(n=3\) to \(\|\mathbf x^{k+1}-\mathbf x^k\|\), \(\|\mathbf z^{k+1}-\mathbf z^k\|\), and \(\|\boldsymbol\lambda^{k+1}-\boldsymbol\lambda^k\|\).\\
For all sufficiently large \(k\), the K{\L} property at
\(\mathbf w^k\) gives
\[
\psi'(e_k)
\operatorname{dist}\!\left(
\mathbf 0,\partial_{\mathbf w}\mathcal P(\mathbf w^k)
\right)
\geq 1.
\]
Combining these inequalities, we obtain
\[
\begin{aligned}
S_{k+1}
&\le
\sqrt{\frac{3}{C_1}}
\left[
\bigl(\psi(e_k)-\psi(e_{k+1})\bigr)
\operatorname{dist}\!\left(\mathbf 0,\partial_\mathbf w\mathcal P(\mathbf w^k)\right)
\right]^\frac{1}{2}.
\end{aligned}
\]
Therefore, for any \(\chi>0\), Young's inequality gives
\[
S_{k+1}
\le
\frac{3\chi}{2C_1}
\bigl(\psi(e_k)-\psi(e_{k+1})\bigr)
+
\frac{1}{2\chi}
\operatorname{dist}\!\left(\mathbf 0,\partial_\mathbf w\mathcal P(\mathbf w^k)\right).
\]
Applying Lemma~\ref{lem:subgradient_bound} at \(\mathbf w^k\) yields
\begin{equation}
S_{k+1}
\le
\frac{3\chi}{2C_1}
\bigl(\psi(e_k)-\psi(e_{k+1})\bigr)
+
\frac{C_2}{2\chi}S_k.
\label{eq:Sk-recursion}
\end{equation}

Choose \(\chi>C_2/2\) and set
\(
\delta:=1-\frac{C_2}{2\chi}>0.
\)
Summing \eqref{eq:Sk-recursion} from \(p=k\) to \(N\), and using the following inequalities:
\[
\begin{aligned}
&\sum_{p=k}^{N}
\bigl(\psi(e_p)-\psi(e_{p+1})\bigr)
\le
\psi(e_k),
&\sum_{p=k}^{N}S_p
\le
S_k+\sum_{p=k}^{N}S_{p+1},
\end{aligned}
\]
we obtain
\[
\delta\sum_{p=k}^{N}S_{p+1}
\le
\frac{3\chi}{2C_1}\psi(e_k)
+
\frac{C_2}{2\chi}S_k.
\]
Letting \(N\to\infty\) gives
\begin{equation}
\begin{aligned}
\sum_{p\ge k}S_{p+1}
&\le
\frac{3\chi}{2C_1\delta}\psi(e_k)+\frac{C_2}{2\chi\delta}S_k.
\end{aligned}
\label{eq:tail-sum-bound}
\end{equation}
Combining \eqref{eq:uk-tail-bound} and
\eqref{eq:tail-sum-bound}, we obtain
\[
\|\mathbf u^k-\mathbf u^\ast\|
\leq
\frac{3\chi}{2C_1\delta}\psi(e_k)
+
\frac{C_2}{2\chi\delta}S_k.
\]
Since
\[
\psi(e_k)
=
\frac{1}{C_3(1-\theta)}e_k^{1-\theta}
\]
and by \eqref{eq:step-error-bound}, we have
\[
S_k
\le
\sqrt{\frac{3}{C_1}}
\sqrt{e_{k-1}-e_k}
\le
\sqrt{\frac{3}{C_1}}\sqrt{e_{k-1}},
\]
it follows that
\[
\|\mathbf u^k-\mathbf u^\ast\|
\leq
C_a e_k^{1-\theta}
+
C_b\sqrt{e_{k-1}},
\]
where
\[
C_a
:=
\frac{3\chi}
{2C_1C_3(1-\theta)\delta}>0,
\quad
C_b
:=
\frac{C_2}{2\chi\delta}
\sqrt{\frac{3}{C_1}}>0.
\]
Since \(\{e_k\}_{k\in\mathbb N}\) is nonincreasing, we have
\(e_k\leq e_{k-1}\), and hence
\(
e_k^{1-\theta}
\leq
e_{k-1}^{1-\theta}.
\)
Therefore,
\begin{equation}
\begin{aligned}
\|\mathbf u^k-\mathbf u^\ast\|
&\leq
C_a e_{k-1}^{1-\theta}
+
C_b\sqrt{e_{k-1}}
\\
&\leq
C\max\left\{
e_{k-1}^{1-\theta},
\sqrt{e_{k-1}}
\right\},
\end{aligned}
\label{eq:sequence-rate-bridge}
\end{equation}
where \(C:=C_a+C_b>0\).

We now distinguish three cases according to the value of \(\theta\).

\paragraph{\(\theta=0\)}
By Theorem~\ref{thm:rate_lyapunov_error}, \(e_k=0\) after finitely many
iterations. It then follows from \eqref{eq:sequence-rate-bridge} that
\(\mathbf u^k=\mathbf u^\ast\) for all sufficiently large \(k\in\mathbb N\).

\paragraph{\(\theta\in(0,1/2]\)}
Since \(1-\theta\ge1/2\) and \(e_k\to0\), for all sufficiently large \(k\in \mathbb N\),
\(
\max\left\{
e_{k-1}^{1-\theta},
\sqrt{e_{k-1}}
\right\}
=
\sqrt{e_{k-1}}.
\)
By~\eqref{eq:linear-rate-ek-explicit}, we get
\[
\|\mathbf u^k-\mathbf u^\ast\|
\le
C\sqrt{e_{k_0}}
\left(
1+\bar C e_{k_0}^{2\theta-1}
\right)^{-\frac{k-1-k_0}{2}},
\quad
\forall k\ge k_0+1.
\]

\paragraph{\(\theta\in(1/2,1)\)}
Since \(1-\theta<1/2\) and \(e_k\to0\), for all sufficiently large \(k\in \mathbb N\),
\(
\max\left\{
e_{k-1}^{1-\theta},
\sqrt{e_{k-1}}
\right\}
=
e_{k-1}^{1-\theta}.
\)
Consequently,~\eqref{eq:sublinear-rate-ek-theorem} gives
\[
\|\mathbf u^k-\mathbf u^\ast\|
\le
C
\left(
\mu(k-1-k_0)+e_{k_0}^{1-2\theta}
\right)^{-\frac{1-\theta}{2\theta-1}}.
\]
This completes the proof.

\section{Proof of Corollary~\ref{cor:eventual-eKKT}}
\label{app:proof_eventual-eKKT}
Define
\[
\begin{aligned}
r_{\mathbf x}^{k+1}
&:=
\operatorname{dist}\left(
\mathbf 0,\,
\nabla F^0(\mathbf x^{k+1})
+\partial F^1(\mathbf x^{k+1})
+\mathbf A^{\mathsf T}\boldsymbol{\lambda}^{k+1}
\right),\\
r_{\mathbf z}^{k+1}
&:=
\operatorname{dist}\left(
\mathbf 0,\,
\nabla H^0(\mathbf z^{k+1})
+\partial H^1(\mathbf z^{k+1})
-\mathbf B^{\mathsf T}\boldsymbol{\lambda}^{k+1}
\right),\\
r_{\mathrm c}^{k+1}
&:=
\left\|
\mathbf A\mathbf x^{k+1}
-\mathbf B\mathbf z^{k+1}
\right\|.
\end{aligned}
\]
The block optimality conditions and the dual update yield
\[
r_{\mathbf x}^{k+1}
\leq
\rho\|\mathbf A^{\mathsf T}\mathbf B\|
\|\mathbf z^{k+1}-\mathbf z^k\|,
\quad
r_{\mathbf z}^{k+1}
\leq
\gamma\|\mathbf Q\|
\|\mathbf z^{k+1}-\mathbf z^k\|.
\]
By Theorem~\ref{thm:rate_iterates}, there exists \(C>0\) such that
\[
\|\mathbf u^k-\mathbf u^\ast\|
\leq
C
\left(
\mu(k-1-k_0)+e_{k_0}^{1-2\theta}
\right)^{-\frac{1-\theta}{2\theta-1}}
\]
for every \(k\geq k_0+1\). Hence, there exist constants \(C_{\mathbf x},C_{\mathbf z}>0\) such that
\[
r_{\mathbf x}^{k+1}\leq C_{\mathbf x}\Phi_k,
\qquad
r_{\mathbf z}^{k+1}\leq C_{\mathbf z}\Phi_k,
\]
where
\[
C_{\mathbf x}=2C\rho\|\mathbf A^{\top}\mathbf B\|,
\quad
C_{\mathbf z}=2C\gamma\|\mathbf Q\|
\]
\[
\Phi_k
:=
\left(
\mu(k-1-k_0)+e_{k_0}^{1-2\theta}
\right)^{-\frac{1-\theta}{2\theta-1}}.
\]
Moreover, since
\(
\mathbf A\mathbf x^\ast-\mathbf B\mathbf z^\ast
=
\frac{\tau}{\rho}\boldsymbol{\lambda}^\ast,
\)
we have
\[
\begin{aligned}
r_{\mathrm c}^{k+1}
&\leq
\epsilon
+\|\mathbf A\|
 \|\mathbf x^{k+1}-\mathbf x^\ast\|
+\|\mathbf B\|
 \|\mathbf z^{k+1}-\mathbf z^\ast\|\\
&\leq
\epsilon+C_{\mathrm c}\Phi_k
\end{aligned}
\]
where \(C_{\mathrm c}=\|\mathbf A\|+\|\mathbf B\|>0\). Therefore, by choosing
\[
C_{\mathrm{KKT}}
\geq
\max\{C_{\mathbf x},C_{\mathbf z},C_{\mathrm c}\},
\]
the three residuals satisfy
\[
\max\left\{
r_{\mathbf x}^{k+1},
r_{\mathbf z}^{k+1},
r_{\mathrm c}^{k+1}
\right\}
\leq
\epsilon_\ast+C_{\mathrm{KKT}}\Phi_k.
\]
For every \(k\geq K_{\bar\epsilon}\ , k\in\mathbb N\), the definition of \(K_{\bar\epsilon}\) gives
\[
\mu(k-1-k_0)+e_{k_0}^{1-2\theta}
\geq
\left(
\frac{C_{\mathrm{KKT}}}
{\bar\epsilon-\epsilon}
\right)^{\frac{2\theta-1}{1-\theta}}.
\]
Consequently,
\[
C_{\mathrm{KKT}}\Phi_k
\leq
\bar\epsilon-\epsilon,
\]
and hence
\[
\max\left\{
r_{\mathbf x}^{k+1},
r_{\mathbf z}^{k+1},
r_{\mathrm c}^{k+1}
\right\}
\leq\bar\epsilon.
\]
Thus, every \(\mathbf u^{k+1}\), \(k\geq K_{\bar\epsilon}\ , k\in\mathbb N\), is a \(\bar\epsilon\)-KKT point.

\section{Proof of Proposition~\ref{prop:inexact-convergence}}
\label{app:proof_inexact_sufficient_descent}
\paragraph{Sufficient descent}

Following the same derivation as in Appendix~\ref{app:proof_combined_descent_rearranged}, the inexact first-order conditions introduce the additional inner-product terms
\[
\left\langle \mathbf e_\mathbf x^{k+1}-\mathbf e_\mathbf x^k,\ \mathbf x^k-\mathbf x^{k+1}\right\rangle
\quad\text{and}\quad
\left\langle \mathbf e_\mathbf z^{k+1}-\mathbf e_\mathbf z^k,\ \mathbf z^k-\mathbf z^{k+1}\right\rangle\\
\]
For any \(\eta_\mathbf x,\eta_\mathbf z>0\), Young's inequality gives
\[
\begin{aligned}
&\left\langle \mathbf e_\mathbf x^{k+1}-\mathbf e_\mathbf x^k,\ \mathbf x^k-\mathbf x^{k+1}\right\rangle\\
&\le
\frac{1}{\eta_\mathbf x}\|\mathbf e_\mathbf x^{k+1}\|^2
+\frac{1}{\eta_\mathbf x}\|\mathbf e_\mathbf x^k\|^2
+\frac{\eta_\mathbf x}{2}\|\mathbf x^{k+1}-\mathbf x^k\|^2 ,
\end{aligned}
\]
\[
\begin{aligned}
&\left\langle \mathbf e_\mathbf z^{k+1}-\mathbf e_\mathbf z^k,\ \mathbf z^k-\mathbf z^{k+1}\right\rangle\\
&\le
\frac{1}{\eta_\mathbf z}\|\mathbf e_\mathbf z^{k+1}\|^2
+\frac{1}{\eta_\mathbf z}\|\mathbf e_\mathbf z^k\|^2
+\frac{\eta_\mathbf z}{2}\|\mathbf z^{k+1}-\mathbf z^k\|^2 .
\end{aligned}
\]
Accordingly, we have the following inexact descent results:
\begin{equation}
\begin{aligned}
&\frac{\rho}{2\eta_1}
\|\mathbf x^{k+1}-\mathbf x^k\|_{\mathbf A^\top\mathbf A}^2
-\mu_{F^0}\|\mathbf x^{k+1}-\mathbf x^k\|^2\\
&
+\|\mathbf z^{k+1}-\mathbf z^k\|_{\mathbf E_\mathbf z}^2
-\frac{\tau}{\rho}
\|\boldsymbol\lambda^{k+1}-\boldsymbol\lambda^k\|^2 \\
&+\frac{\eta_\mathbf x}{2}\|\mathbf x^{k+1}-\mathbf x^k\|^2
+\frac{\eta_\mathbf z}{2}\|\mathbf z^{k+1}-\mathbf z^k\|^2\\
&+\frac{1}{\eta_\mathbf x}\|\mathbf e_\mathbf x^{k+1}\|^2
+\frac{1}{\eta_\mathbf x}\|\mathbf e_\mathbf x^{k}\|^2
+\frac{1}{\eta_\mathbf z}\|\mathbf e_\mathbf z^{k+1}\|^2
+\frac{1}{\eta_\mathbf z}\|\mathbf e_\mathbf z^{k}\|^2\\
\ge&
\|\mathbf z^{k+1}-\mathbf z^k\|_{\mathbf D_\mathbf z}^2
-\|\mathbf z^k-\mathbf z^{k-1}\|_{\mathbf D_\mathbf z}^2 \\
&
+\frac{1-\tau}{2\rho}
\|\boldsymbol\lambda^{k+1}-\boldsymbol\lambda^k\|^2
-\frac{1-\tau}{2\rho}
\|\boldsymbol\lambda^k-\boldsymbol\lambda^{k-1}\|^2 .
\end{aligned}
\label{eq:combined-descent-rearranged-inexact}
\end{equation}

Similarly, the descent arguments in Appendix~\ref{app:proof_one_step_descent_AL} extend to the inexact updates by accounting for the additional residual term introduced in the first-order condition of each primal block. Applying Young's inequality to these terms gives, respectively,
\begin{equation}
\begin{aligned}
&\mathcal L_{\rho,\tau}(\mathbf x^{k+1},\mathbf z^k,\boldsymbol\lambda^k)-\mathcal L_{\rho,\tau}(\mathbf x^k,\mathbf z^k,\boldsymbol\lambda^k) \\
&\le
-\frac{1}{2}
\|\mathbf x^{k+1}-\mathbf x^k\|^2_{\mu_{F^0}\mathbf I+\rho\mathbf A^\top\mathbf A}\\
&\qquad 
+\frac{1}{2\alpha_\mathbf x}\|\mathbf e_\mathbf x^{k+1}\|^2+\frac{\alpha_\mathbf x}{2}\|\mathbf x^{k+1}-\mathbf x^k\|^2.
\end{aligned}
\label{eq:ALM-x-update-descent-FOC}
\end{equation}
\begin{equation}
\begin{aligned}
&\mathcal L_{\rho,\tau}(\mathbf x^{k+1},\mathbf z^{k+1},\boldsymbol\lambda^k)
-\mathcal L_{\rho,\tau}(\mathbf x^{k+1},\mathbf z^k,\boldsymbol\lambda^k) \\
&\leq 
-\frac{1}{2}\|\mathbf z^{k+1}-\mathbf z^k\|^2_{
(\rho-\omega_{H^0})\mathbf B^\top \mathbf B+2\gamma \mathbf Q}\\
&\qquad 
+\frac{1}{2\alpha_\mathbf z}\|\mathbf e_\mathbf z^{k+1}\|^2+\frac{\alpha_\mathbf z}{2}\|\mathbf z^{k+1}-\mathbf z^k\|^2.
\end{aligned}
\label{eq:ALM-z-update-descent-FOC}
\end{equation}
Combining~\eqref{eq:ALM-lambda-update} and \eqref{eq:combined-descent-rearranged-inexact}-\eqref{eq:ALM-z-update-descent-FOC}, we obtain
\begin{equation}
\begin{aligned}
&\mathcal P(\mathbf w^{k+1})-\mathcal P(\mathbf w^k) \\
\le&
-\frac{1}{2}
\left\|\mathbf x^{k+1}-\mathbf x^k\right\|^2_{\mathbf M_\mathbf x}
-\frac{1}{2}
\left\|\mathbf z^{k+1}-\mathbf z^k\right\|^2_{\mathbf M_\mathbf z}\\
&-a_{\boldsymbol\lambda}
\|\boldsymbol\lambda^{k+1}-\boldsymbol\lambda^k\|^2 \\
&+d\frac{\eta_\mathbf x}{2}\|\mathbf x^{k+1}-\mathbf x^k\|^2
+d\frac{\eta_\mathbf z}{2}\|\mathbf z^{k+1}-\mathbf z^k\|^2\\
&+d\frac{1}{\eta_\mathbf x}\|\mathbf e_\mathbf x^{k+1}\|^2
+d\frac{1}{\eta_\mathbf x}\|\mathbf e_\mathbf x^{k}\|^2
+d\frac{1}{\eta_\mathbf z}\|\mathbf e_\mathbf z^{k+1}\|^2
+d\frac{1}{\eta_\mathbf z}\|\mathbf e_\mathbf z^{k}\|^2\\
&
+\frac{1}{2\alpha_\mathbf x}\|\mathbf e_\mathbf x^{k+1}\|^2+\frac{\alpha_\mathbf x}{2}\|\mathbf x^{k+1}-\mathbf x^k\|^2\\
&
+\frac{1}{2\alpha_\mathbf z}\|\mathbf e_\mathbf z^{k+1}\|^2+\frac{\alpha_\mathbf z}{2}\|\mathbf z^{k+1}-\mathbf z^k\|^2\\
\le&
-\frac{1}{2}
\left\|\mathbf x^{k+1}-\mathbf x^k\right\|^2_{\mathbf{\widetilde{M}_\mathbf x}}
-\frac{1}{2}
\left\|\mathbf z^{k+1}-\mathbf z^k\right\|^2_{\mathbf{\widetilde{M}_\mathbf z}}\\
&-a_{\boldsymbol\lambda}
\|\boldsymbol\lambda^{k+1}-\boldsymbol\lambda^k\|^2+\varepsilon_{k+1}\\
\end{aligned}
\label{eq:Lyapunov-descent-combined-inexact}
\end{equation}
where
\[
\begin{aligned}
&\mathbf{\widetilde{M}_x}:= \mathbf M_\mathbf x-\left(d\eta_\mathbf x+\alpha_\mathbf x\right)\mathbf I,\\
&\mathbf{\widetilde{M}_z}:= \mathbf M_\mathbf z-\left(d\eta_\mathbf z+\alpha_\mathbf z\right)\mathbf I,\\
&a_{\boldsymbol\lambda}:=
\frac{d\tau}{\rho}
-\frac{(1-\tau)(2-\tau)}{2\rho}>0,\\
&\varepsilon_{k+1}:=\left(d\frac{1}{\eta_\mathbf x}+\frac{1}{2\alpha_\mathbf x}\right)\|\mathbf e_\mathbf x^{k+1}\|^2
+d\frac{1}{\eta_\mathbf x}\|\mathbf e_\mathbf x^{k}\|^2\\
&\qquad
+\left(d\frac{1}{\eta_\mathbf z}+\frac{1}{2\alpha_\mathbf z}\right)\|\mathbf e_\mathbf z^{k+1}\|^2
+d\frac{1}{\eta_\mathbf z}\|\mathbf e_\mathbf z^{k}\|^2\geq0.
\end{aligned}
\]
Since \(\mathbf M_\mathbf x\succ\mathbf 0\) and \(\mathbf M_\mathbf z\succ\mathbf 0\), the auxiliary parameters \(\eta_\mathbf x,\eta_\mathbf z,\alpha_\mathbf x,\alpha_\mathbf z>0\) can be chosen sufficiently small so that
\[
d\eta_\mathbf x+\alpha_\mathbf x<\lambda_{\min}(\mathbf M_\mathbf x),
\qquad
d\eta_\mathbf z+\alpha_\mathbf z<\lambda_{\min}(\mathbf M_\mathbf z).
\]
Consequently,
\(\widetilde{\mathbf M}_\mathbf x\succ\mathbf 0\) and \(\widetilde{\mathbf M}_\mathbf z\succ\mathbf 0\).
Therefore, Dropping the nonpositive terms in~\eqref{eq:Lyapunov-descent-combined-inexact} and summing from \(j=0\) to \(k\) gives
\[
\begin{aligned}
\mathcal P(\mathbf w^{k+1})
&\leq
\mathcal P(\mathbf w^0)
+
\sum_{j=0}^{k}\varepsilon_{j+1}\\
&\leq
\overline{\mathcal P}
:=
\mathcal P(\mathbf w^0)
+
\sum_{j=0}^{\infty}\varepsilon_{j+1},
\qquad
\forall\,k\in\mathbb N.
\end{aligned}
\]
Thus, the inexact Lyapunov sequence is uniformly bounded from above.
\paragraph{Lower boundedness}
Unlike the exact case, the Lyapunov sequence \(\{\mathcal P(\mathbf w^k)\}_{k\in\mathbb N}\) is not necessarily nonincreasing. Since \(\varepsilon_{k+1}\) is a nonnegative linear combination of
\(\|\mathbf e_{\mathbf x}^{k+1}\|^2\),
\(\|\mathbf e_{\mathbf x}^{k}\|^2\),
\(\|\mathbf e_{\mathbf z}^{k+1}\|^2\), and
\(\|\mathbf e_{\mathbf z}^{k}\|^2\), the summability condition in~\eqref{eq:summable-subproblem-errors} implies that
\(
\sum_{k=0}^{\infty}\varepsilon_{k+1}<+\infty.
\)
Therefore, dropping the nonpositive terms in~\eqref{eq:Lyapunov-descent-combined-inexact} and summing from \(j=0\) to \(k\) yield
\[
\begin{aligned}
\mathcal P(\mathbf w^{k+1})
&\leq
\mathcal P(\mathbf w^0)
+
\sum_{j=0}^{k}\varepsilon_{j+1}\\
&\leq
\overline{\mathcal P}
:=
\mathcal P(\mathbf w^0)
+
\sum_{j=0}^{\infty}\varepsilon_{j+1}<+\infty,
\qquad
\forall\,k\in\mathbb N.
\end{aligned}
\]
Thus, the inexact Lyapunov sequence is uniformly bounded from above. The remainder of the boundedness argument follows as in the exact case, with \(\mathcal P(\mathbf w^0)\) replaced by
\(\overline{\mathcal P}\).
Accordingly, define
\[
M
:=
\max\left\{
\|\boldsymbol\lambda^0\|^2,
\frac{2\rho}{\tau(1-\tau)}
\left(\overline{\mathcal P}-\underline{\mathcal P}\right)
\right\}.
\]
The induction argument then yields
\(
\|\boldsymbol\lambda^k\|^2\leq M,
\quad
\forall\,k\in\mathbb N.
\)
Consequently, with this modified bound, \(\{\mathcal P(\mathbf w^k)\}_{k\in\mathbb N}\) is bounded from below.

\paragraph{Stationarity and \(\epsilon\)-KKT Conditions}

From the inexact first-order conditions of the two block subproblems
and the dual update, we obtain
\[
\begin{aligned}
&\operatorname{dist}\!\left(
\mathbf 0,
\nabla F^0(\mathbf x^{k+1})
+\partial F^1(\mathbf x^{k+1})
+\mathbf A^\top\boldsymbol\lambda^{k+1}
\right)\\
&\leq
\rho\|\mathbf A^\top\mathbf B\|
\|\mathbf z^{k+1}-\mathbf z^k\|
+
\|\mathbf e_\mathbf x^{k+1}\|,
\end{aligned}
\]
and
\[
\begin{aligned}
&\operatorname{dist}\!\left(
\mathbf 0,
\nabla H^0(\mathbf z^{k+1})
+\partial H^1(\mathbf z^{k+1})
-\mathbf B^\top\boldsymbol\lambda^{k+1}
\right)\\
&\leq
\gamma\|\mathbf Q\|
\|\mathbf z^{k+1}-\mathbf z^k\|
+
\|\mathbf e_\mathbf z^{k+1}\|.
\end{aligned}
\]
By~\eqref{eq:Lyapunov-descent-combined-inexact}, the lower boundedness of the
Lyapunov sequence, and the summability of
\(\{\varepsilon_{k+1}\}_{k\in\mathbb N}\), the successive differences
are square summable. Hence,
\(
\|\mathbf z^{k+1}-\mathbf z^k\|\to0.
\)
In addition, the assumed square summability of the inexactness
residuals gives
\(
\|\mathbf e_{\mathbf x}^{k+1}\|\to0,
\
\|\mathbf e_{\mathbf z}^{k+1}\|\to0.
\)
Hence, both stationarity residuals converge to zero. Passing to the limit along a convergent
subsequence shows that
\[
\mathbf 0
\in
\nabla F^0(\mathbf x^\ast)
+\partial F^1(\mathbf x^\ast)
+\mathbf A^\top\boldsymbol\lambda^\ast
\]
and
\[
\mathbf 0
\in
\nabla H^0(\mathbf z^\ast)
+\partial H^1(\mathbf z^\ast)
-\mathbf B^\top\boldsymbol\lambda^\ast.
\]
Together with the same asymptotic feasibility bound as in the exact case, this shows that every accumulation point of the inexact iterate sequence is an \(\epsilon\)-KKT point.

\section{Proof of Definability of \(\mathcal P\) in \(\mathbb R_{\exp}\)}
\label{app:lyapunov_definability}
Recall that, for a definable set \(A\subseteq\mathbb R^n\), a function
\(f:A\to\mathbb R^m\) is definable if its graph
\(
\operatorname{gph} f
:=
\{(\mathbf x,f(\mathbf x)):\mathbf x\in A\}
\)
is an \(\mathbb R_{\exp}\)-definable subset of \(\mathbb R^{n+m}\)~\cite{van1998tame}.
We shall use the following standard properties of definable sets and functions~\cite{van1996geometric}. \\
i) Every semialgebraic set is \(\mathbb R_{\exp}\)-definable, since \(\mathbb R_{\exp}\) expands the real field. \\
ii) If \(f_i:\mathbb R^m\to\mathbb R\), \(i=1,\ldots,M\), are \(\mathbb R_{\exp}\)-definable functions, then any finite sum, product, scalar multiple, and composition of the \(f_i\)'s is again \(\mathbb R_{\exp}\)-definable. \\
iii) If \(C\subset\mathbb R^m\) is \(\mathbb R_{\exp}\)-definable and \(f:\mathbb R^m\to\mathbb R\) is \(\mathbb R_{\exp}\)-definable, then the sets
\[
\begin{aligned}
\{u&\in C:f(u)=0\},
\quad
\{u\in C:f(u)\ge 0\},\\
\quad
\{(u,r):u&\in C,\ r=f(u)\},
\quad
\{(u,r):u\in C,\ r\ge f(u)\}
\end{aligned}
\]
are \(\mathbb R_{\exp}\)-definable. These facts follow from the closure of definable sets under finite Boolean operations, Cartesian products, and projections.
We now verify the \(\mathbb R_{\exp}\)-definability of each component of \(\mathcal{P}\).

\paragraph{Definability of \(F^1(\mathbf x)=\iota_{\Delta_{\mathcal U}}\) and \(F^1(\mathbf z)=\iota_{\mathcal Q}\)}~{}
\newline
The  probability simplex set \(\Delta_{\mathcal U}\) is a polyhedral. Since polyhedral sets are semialgebraic, \(\Delta_{\mathcal U}\) is \(\mathbb R_{\exp}\)-definable.
The finite-valued graph of \(\iota_{\Delta_{\mathcal U}}\) is
\[
\operatorname{gph}\iota_{\Delta_{\mathcal U}}
=
\{(\mathbf x,r)\in\mathbb R^{n_\mathbf x}\times\mathbb R:
\mathbf x\in\Delta_{\mathcal U},\ r=0\}.
\]
Since both conditions \(\mathbf x\in\Delta_{\mathcal U}\) and \(r=0\) are \(\mathbb R_{\exp}\)-definable, \(\operatorname{gph}\iota_{\Delta_{\mathcal U}}\) is \(\mathbb R_{\exp}\)-definable. Thus \(\iota_{\Delta_{\mathcal U}}\) is an \(\mathbb R_{\exp}\)-definable extended-valued function.
Equivalently,
\[
\operatorname{epi}\iota_{\Delta_{\mathcal U}}
=
\{(\mathbf x,r):\mathbf x\in\Delta_{\mathcal U},\ r\ge0\}
=
\Delta_{\mathcal U}\times[0,+\infty)
\]
is \(\mathbb R_{\exp}\)-definable.
Similarly,
\[
\operatorname{epi}\iota_{\mathcal Q}
=
\{(\mathbf z,r):\mathbf z\in\mathcal Q,\ r\ge0\}
=
\mathcal Q\times[0,+\infty)
\]
is \(\mathbb R_{\exp}\)-definable.

\paragraph{Definability of the entropy-related terms}~{}
\newline
Define
\[
h(t):=
\begin{cases}
t\log t, & t>0,\\
0, & t=0.
\end{cases}
\]
The logarithm is \(\mathbb R_{\exp}\)-definable. Indeed, by the graph criterion for definable maps,
\[
\begin{aligned}
\operatorname{gph}(\log)
&=
\{(t,y)\in\mathbb R^2:t>0,\ y=\log t\}\\
&=
\{(t,y)\in\mathbb R^2:t>0,\ t=\exp(y)\},
\end{aligned}
\]
which is \(\mathbb R_{\exp}\)-definable because \(\exp\), the order relation, and equality are definable in \(\mathbb R_{\exp}\). Since multiplication is also definable in \(\mathbb R_{\exp}\), the function
\(
t\mapsto t\log t
\)
is \(\mathbb R_{\exp}\)-definable on \((0,+\infty)\). Furthermore, \(h\) is obtained from \(t\mapsto t\log t\) and the constant function \(0\) by a finite piecewise definable construction on the definable sets \((0,+\infty)\) and \(\{0\}\). Hence \(h\) is \(\mathbb R_{\exp}\)-definable on \([0,+\infty)\).

For finite alphabets, entropy and conditional entropy can be written as finite sums of functions of the form
\(
h(a_i(\cdot)),
\)
where each \(a_i(\cdot)\) is an affine probability expression or a marginal probability obtained by finite summation. Since each \(a_i\) is an affine map in finitely many variables, it is polynomial and therefore \(\mathbb R_{\exp}\)-definable. By closure of definable functions under composition, each \(h\circ a_i\) is \(\mathbb R_{\exp}\)-definable. By closure under finite sums and scalar multiplications, it follows that \(H(U),H(U|X),H(U|Y)\) are \(\mathbb R_{\exp}\)-definable. Consequently, \(F^0(\mathbf x),H^0(\mathbf z)\) are \(\mathbb R_{\exp}\)-definable.

\paragraph{Definability of \(\langle \boldsymbol\lambda,\mathbf A\mathbf x-\mathbf B\mathbf z\rangle\)}~{}
\newline
For fixed matrices \(\mathbf A\) and \(\mathbf B\), this term is
\[
\langle \boldsymbol\lambda,\mathbf A\mathbf x-\mathbf B\mathbf z\rangle
=
\boldsymbol\lambda^\top \mathbf A\mathbf x-\boldsymbol\lambda^\top \mathbf B\mathbf z.
\]
In coordinates, this is a finite sum of monomials of degree at most two in the entries of \((\mathbf x,\mathbf z,\boldsymbol\lambda)\). Therefore, it is a polynomial function
\[
(\mathbf x,\mathbf z,\boldsymbol\lambda)\mapsto \boldsymbol\lambda^\top \mathbf A\mathbf x-\boldsymbol\lambda^\top \mathbf B\mathbf z
\]
on a finite-dimensional Euclidean space. Hence it is semialgebraic and thus \(\mathbb R_{\exp}\)-definable.

\paragraph{Definability of quadratic and norm terms}~{}
\newline
The quadratic penalty term satisfies
\[
\|\mathbf A\mathbf x-\mathbf B\mathbf z\|^2
=
(\mathbf A\mathbf x-\mathbf B\mathbf z)^\top(\mathbf A\mathbf x-\mathbf B\mathbf z),
\]
which is a polynomial function of \((\mathbf x,\mathbf z)\). Hence
\[
(\mathbf x,\mathbf z)\mapsto \frac{\rho}{2}\|\mathbf A\mathbf x-\mathbf B\mathbf z\|^2
\]
is \(\mathbb R_{\exp}\)-definable.
Similarly, the Lyapunov correction terms satisfy
\[
\|\mathbf B\mathbf z-\mathbf B\bar{\mathbf z}\|^2
=
(\mathbf B\mathbf z-\mathbf B\bar{\mathbf z})^\top(\mathbf B\mathbf z-\mathbf B\bar{\mathbf z}),
\]
\[
\|\boldsymbol\lambda-\bar{\boldsymbol\lambda}\|^2
=
(\boldsymbol\lambda-\bar{\boldsymbol\lambda})^\top(\boldsymbol\lambda-\bar{\boldsymbol\lambda}),
\]
and
\[
\|\mathbf z-\bar{\mathbf z}\|_{\mathbf Q}^2
=
(\mathbf z-\bar{\mathbf z})^\top \mathbf Q(\mathbf z-\bar{\mathbf z}).
\]
Since \(\mathbf B\) and \(\mathbf Q\) are fixed matrices, each of these terms is a polynomial function.\\
Moreover,
\(
\|\boldsymbol\lambda\|^2=\boldsymbol\lambda^\top\boldsymbol\lambda
\)
is polynomial in \(\boldsymbol\lambda\). Therefore all quadratic and norm terms in \(\mathcal P\) are semialgebraic and hence \(\mathbb R_{\exp}\)-definable.

\section{Properness and Lower Semicontinuity of \(\mathcal P\)}
\label{app:proper_lsc_P}

Recall that
\(
\mathbf w
=
\bigl(
\mathbf x,
\mathbf z,
\boldsymbol\lambda,
\mathbf z',
\boldsymbol\lambda'
\bigr).
\)
Define
\[
\mathcal C
:=
\Delta_{\mathcal U}
\times
\mathcal Q
\times
\mathbb R^{n_{\boldsymbol\lambda}}
\times
\mathcal Q
\times
\mathbb R^{n_{\boldsymbol\lambda}}.
\]
Then, the Lyapunov function admits the representation
\[
\mathcal P(\mathbf w)
=
\Psi(\mathbf w)+\iota_{\mathcal C}(\mathbf w),
\]
where \(\Psi\) comprises all real-valued terms of \(\mathcal P\), and
\[
\iota_{\mathcal C}(\mathbf w)
=
\iota_{\Delta_{\mathcal U}}(\mathbf x)
+
\iota_{\mathcal Q}(\mathbf z).
\]
Since \(\Delta_{\mathcal U}\) and \(\mathcal Q\) are nonempty, \(\mathcal C\neq\varnothing\). Under the convention \(0\log 0=0\), the entropy terms in \(\Psi\) are finite on the corresponding probability simplices, and all remaining terms are finite-valued. Hence,
\(
\operatorname{dom}\mathcal P
=
\mathcal C
\neq
\varnothing.
\)
Furthermore,
\[
\mathcal P(\mathbf w)>-\infty,
\qquad
\forall\,\mathbf w.
\]
Therefore, \(\mathcal P\) is proper. To establish lower semicontinuity, recall that the indicator function \(\iota_{\mathcal C}\) is lower semicontinuous if and only if its set \(\mathcal C\) is closed~\cite[Chapter~1]{rockafellar1998variational}. Since \(\Delta_{\mathcal U}\) and \(\mathcal Q\) are closed, their Cartesian product with the corresponding Euclidean spaces is also closed. Therefore, \(\mathcal C\) is closed, and hence \(\iota_{\mathcal C}\) is lower semicontinuous. Moreover, the finite-valued part \(\Psi\) is continuous on \(\mathcal{C}\). Hence, as the sum of a continuous finite-valued function and lower semicontinuous extended-valued functions, \(\mathcal P\) is lower semicontinuous.

\bibliographystyle{IEEEtran}
\bibliography{refs}
\newpage

\end{document}